# TRUE MASSES OF RADIAL-VELOCITY EXOPLANETS

Robert A. Brown

rbrown@stsci.edu

Space Telescope Science Institute

January 12, 2014

**ABSTRACT**

We explore the science power of space telescopes used to estimate the true masses of known radial-velocity exoplanets by means of astrometry on direct images. We translate a desired mass accuracy—±10% in our example—into a minimum goal for the signal-to-noise ratio, which implies a minimum exposure time. When the planet is near a node, the mass measurement becomes difficult if not impossible, because the apparent separation becomes decoupled from the inclination angle of the orbit. The combination of this nodal effect with considerations of solar and anti-solar pointing restrictions, photometric and obscurational completeness, and image blurring due to orbital motion, severely limits the observing opportunities, often to only brief intervals in a five-year mission. We compare the science power of four missions, two with external star shades, *EXO-S and WFIRST-S*, and two with internal coronagraphs, *EXO-C* and *WFIRST-C*. The star shades out-perform the coronagraph in this science program by about a factor of three. For both coronagraphs, the input catalog includes 16 RV planets, of which *EXO-C* could possibly observe 10, of which 6 would have a 90% guarantee of success. Of the same 16 planets, *WFIRST-C* could possibly observe 12, of which 9 are guaranteed. For both star-shade missions, the input catalog includes 55 planets, of which *EXO-S* could possibly observe 37, of which 20 are guaranteed. Of the same 55, *WFIRST-S* could possibly observe 45, of which 30 are guaranteed. The longer spectroscopic exposure times should be easily accommodated for the RV planets with guaranteed success.

*Subject headings*: methods: statistical, planetary systems, planets and satellites: detection,

## 1. INTRODUCTION

To date, hundreds of exoplanets have been discovered by the radial-velocity (RV) technique. Not only have these objects changed the course of the history of astronomy and opened rich new fields of study, the known RV planets remain special objects of further research, mainly because they are relatively nearby, and because they revolve around bright, well-known stars. Without doubt, scientists will one day know each of these RV planets intimately, as individual objects, each with unique origins, attributes and environments, just like the solar-system planets today. Nevertheless, we can now only charter the next steps toward that goal, which are to determine the true masses ($m_p$) of known RV planets and to obtain simple diagnostic spectra of their atmospheres— amounting to the initial characterization of known RV planets. Towards that end, this paper sets down a theory for the tasks of mass determination and spectral characterization of RV planets, and it uses that theory to estimate the performance of four space optical systems currently under study. These missions have been motivated, in part, by the emerging opportunities to directly image and characterize RV planets.

Why are "theories" needed in this context? For science-operational reasons, to be clear and distinct about how we structure and optimize the observing program, analyze the data, and about how to estimate the science outcomes for alternative mission designs.

In the case of spectral characterization, the theory is simple. The plan is to obtain a spectrum—resolving power $\mathcal{R} = 70$ in *I* band and signal-to-noise ratio S/N = 10 in the continuum—*immediately* after directly imaging a possible RV planet. The reason for acting immediately is that the source may shortly disappear or otherwise become unobservable.

In the case of mass determination, the theory we offer is new, complicated, and interesting. Indeed, to the best of our knowledge, no science requirements on mass-accuracy have yet been proposed in print, nor have any measurement requirements been derived from any such requirements, nor has the performance of candidate missions so far been tested on the sample of known RV planets taking into account all key factors. This paper provides the needed theory, which entails measurement requirements that depend strongly on the orbital phase of the planet, especially near node crossings. The treatment also involves solar-avoidance restrictions, variable obscuration near the central field stop, a systematic limit to the depth of search exposures, and the blurring of images due to orbital motion. The treatment of Traub (2014) takes none of these factors into account.

As is well known, RV observations estimate the *projected* mass of an exoplanet, $m_p \sin i$, where *i* is the usually unknown inclination of the planetary orbit. Today, we know $m_p \sin i$ for hundreds of RV planets, but we know $m_p$ only for a few: the transiting exoplanets with known RV orbits, for which $\sin i \approx 1$, and $m_p$ is trivial to compute. For non-transiting RV planets, we must make some new measurement that constrains *i* and therefore removes the "sin *i* degeneracy" in planetary mass. Here, there are at least two possibilities. Brown (2009b) presents a theory for estimating *i* from the "photometric orbits" of reflected starlight from distant, non-transiting, *Kepler*-type exoplanets with RV orbital solutions. In this paper, we present a complementary theory, which is applicable to nearby RV planets. It estimates *i* from the apparent separation (*s*) between the planet and the host star, where *s* is measured by direct astrometry. In this case, a single measurement of *s* is, in principle, sufficient to estimate *i*. However, both the absolute time and the exposure time are critical to the feasibility and accuracy of estimating $m_p$.

Although we are not aware of any other way to convert *individual* values of $m_p \sin i$ into values of $m_p$—other than the theory in this paper—various statistical approaches have been offered for estimating the *distribution* of a "projected" quantity (i.e. a quantity with a factor sin *i*), including rotation rates of unresolved stars (Chandrasekhar & Münch 1950); microlensing (Giannini & Lunine 2013; Brown 2014); and RV planets (Jorisson, A., Mayor & Udry 2001; Zucker, & Mazeh 2001; Marcy, G. W., et al. 2005; Udry & Santos 2007; Brown 2011; Ho & Turner 2011; Lopez & Jenkins 2012). Nevertheless, Brown (2011) shows that a usefully constrained distribution of an unprojected quantity may demand a sample of thousands or tens of thousands projected values, certainly more than the mere hundreds we have in the case of currently known RV planets.

Two lines of research motivate our working science requirement for mass determination. The first comprises studies of the formation and evolution of planetary systems. Just as explaining the distributions of planetary mass and orbital distance have long been a focus of solar-system studies, today such research has been taken to the next level, addressing those distributions broadly—ultimately, everywhere in the universe. The second type comprises studies of planetary atmospheres, where the planetary mass enters into the scale height. Knowing the scale height is important for interpreting planetary spectra, which is relevant because we plan follow-on spectroscopy as an adjunct to the mass-determination program. Informal discussions with scientists involved in these missions suggest that a useful science requirement would be: for as many radial-velocity planets as possible, obtain the planetary masses to within ± 10% uncertainty:

$$\frac{\sigma_{m_p}}{m_p} = 0.1 \tag{1}$$

where $\sigma_{m_p}$ is the standard deviation of $m_p$. In this paper, we adopt Eq. (1) as a baseline requirement.

§2 presents our theory for estimating $m_p$ with a required accuracy $\sigma_{m_p}$. §3 is *mise en place*, describing three basics: in §3.1, the baseline parameters and assumptions for the four missions under study; in §3.2, the input catalog of potentially observable RV planets; and in §3.3, the exposure time calculator. §4 defines the *cube of intermediate information* that realizes a mission and can generate various diagnostic reports. §5 introduces the concept of the design reference mission (DRM), including scheduling algorithm, merit function, union completeness, and median exposure times. §6 compares the baseline performance of four missions that share the goal of determining $m_p$ for known RV planets. These missions, which could be implemented in the timeframe of the 2020s, are *WFIRST-C*, a 2.4 m telescope with an internal coronagraph; *WFIRST-S*, a 2.4 m telescope with an external star shade; *EXO-S*, a 1.1 m telescope with a star shade, and *EXO-C*, a 1.4 m telescope with an internal coronagraph. §7 interprets and concludes.

## 2. THEORY OF $\sigma_{m_p}/m_p$

In this section, we deconstruct the desired fractional accuracy in $m_p$ to determine the S/N required by Equation (1). In §2.1, in the first step in this deconstruction, we find the required astrometric error, $\sigma_s$, which depends strongly on the orbital phase of the planet. In §2.2, the second step involves details of the "roll convolution" protocol (§2.2.1) and an empirical calibration of $\sigma_s$ as a function S/N (§2.2.2).

At that point, knowing in principle how to find the value of S/N that satisfies Equation (1)—for any design parameters, RV planet, Julian day, and random value of inclination—we can use the basics described in §3 to compute exposure times ($\tau$), populate the cube of intermediate information described in §4, and produce design reference missions (DRMs) as described in §5 and exemplified in §6.

### 2.1. REQUIRED ASTROMETRIC UNCERTAINTY, $\sigma_s$

The seven elements of the planetary orbit fall in three groups. Semimajor axis $a$ and eccentricity $e$ are intrinsic to the true orbital ellipse in space. The argument of periapsis $\omega_p$, the inclination $i$, and the longitude of the ascending node $\Omega$ are three ordered rotations that define the *apparent* orbit on the plane of the sky. The period $P$, the current Julian day $t$, and the Julian day of any periapsis passage $t_0$ determine the orbital phase, which is the mean anomaly $M$, according to Eq. (6), below.

RV observations provide values for five of the seven orbital elements: $a$, $e$, $P$, $t_0$, and $\omega_s$, where $\omega_s$ is the argument of periapsis *of the star*, from which we can compute $\omega_p = \omega_s + 180°$ *for the planet*. In this preliminary treatment, we assume the orbital elements are without error. The delay from the time this paper is being composed until the start of one of these missions—allowing at least ten more years of RV observations—should allow this goal to be achieved for practical purposes.

We introduce a Cartesian coordinate system with the star at the origin, $+x$ toward north, $+y$ toward east, and $+z$ toward the observer. The base $a$-$e$ orbit lies in the plane of the sky through the star ($z = 0$) with the semimajor axis on the $x$ axis and the periapsis on $+x$. The $x$-$y$ position of the planet in the base orbit is described by the parametric equations

$$\begin{aligned} x_0 &= r(\nu)\cos\nu \\ y_0 &= r(\nu)\sin\nu \end{aligned}, \qquad (2)$$

where the planet-star distance is

$$r(\nu) = \frac{a(1-e^2)}{1+e\cos\nu}, \qquad (3)$$

with $r$ in arcseconds for $a$ in arcseconds, where $v$ is the true anomaly, which is the root of the equation

$$\tan\frac{v}{2} = \sqrt{\frac{1+e}{1-e}} \tan\frac{E}{2}, \tag{4}$$

in which the eccentric anomaly ($E$) is the root of Kepler's Equation,

$$E - e\sin E = M, \tag{5}$$

in which the mean anomaly or orbital phase is:

$$M = 2\pi\left(\frac{t-t_0}{P}\right). \tag{6}$$

The true position of the planet in space, $(x, y, z)$, is found by first rotating the vector $(x_0, y_0, 0)$ by $\omega_p$ around the $+z$ axis, then by $i$ around the $-y$ axis, and lastly by $\Omega$ around the $+z$ axis.

Setting $\Omega = 0$ because it is irrelevant to this study, the apparent position of the planet is

$$\begin{aligned} x &= x_0 \cos\omega_p - y_0 \sin\omega_p \\ y &= \left(x_0 \sin\omega_p + y_0 \cos\omega_p\right)\cos i \end{aligned} \tag{7}$$

The apparent separation in arcseconds between star and planet is

$$s = \sqrt{\left(x_0 \cos\omega_p - y_0 \sin\omega_p\right)^2 + \left(x_0 \sin\omega_p + y_0 \cos\omega_p\right)^2 \cos^2 i}. \tag{8}$$

If we were analyzing an *actual* image of an RV planet, we would estimate $m_p$ by dividing the value of $m_p \sin i$ by the value of $\sin i$ inferred from the planetary image. The actual measured quantity in such an image would be the apparent separation $s$—not $i$ or $\sin i$ itself. To obtain $\sin i$ from $s$, we invert Equation (8):

$$\sin i = \frac{\sqrt{x_0^2 + y_0^2 - s^2}}{\left|x_0 \sin\omega_p + y_0 \cos\omega_p\right|}. \tag{9}$$

We see from Equations (8–9) that only a limited range of $s$ is possible for any radial-velocity planet, $s_{min} \leq s \leq s_{max}$, where

$$s_{min} = \left|x_0 \cos\omega_p - y_0 \sin\omega_p\right| \tag{10}$$

and

$$s_{max} = \sqrt{x_0^2 + y_0^2},\tag{11}$$

which is the true, physical separation in space. Here, all distances are rendered in arcseconds.

To compute the standard deviation (astrometric error) of $s$ ($\sigma_s$), we must approach the foregoing information from a different direction, starting with $\sin i$, for which we know the distribution but not the value. The *uncertainty* in $\sin i$ ($\sigma_{\sin i}$) is controlled by the science requirement on mass accuracy: $\sigma_{\sin i} = \sin i\,(\sigma_{m_p}/m_p)$, where $\sigma_{m_p}$ is the desired uncertainty in $m_p$. Therefore, from Equation (9), in the linear approximation,

$$\begin{aligned}\sigma_s &= \sigma_{\sin i}\left|\frac{\partial \sin i}{\partial s}\right|^{-1}\\ &= \sigma_{\sin i}\left|\frac{x_0 \sin\omega_p + y_0 \cos\omega_p \sqrt{x_0^2 + y_0^2 - s^2}}{s}\right|\\ &= \frac{x_0^2 + y_0^2 - s^2}{s}\frac{\sigma_{m_p}}{m_p}\end{aligned}\tag{12}$$

It is apparent from Equation (12) that $\sigma_s$ goes to zero as $s$ goes to $\sqrt{x_0^2 + y_0^2}$, which occurs when the planet passes through a node and crosses the plane of the sky. At this time, the sensitivity of $s$ to $i$ vanishes, and therefore it temporarily becomes impossible to estimate $i$ from a measurement of $s$, no matter how precise.

Table 1 gives all the node passages during the baseline five-year mission for the 55 RV planets in play for the missions studied in this paper. (The input catalog of known RV planets is discussed in §3.2.)

Table 1. Node crossings for known RV planets in play in 2024–2029.

| star name | HIP | mission day | Julian day | date |
|---|---|---|---|---|
| GJ 317 b | … | 34.15 | 2460344.65 | 04 February 2024 |
| | | 356.73 | 2460667.23 | 22 December 2024 |
| | | 727.05 | 2461037.55 | 28 December 2025 |
| | | 1049.63 | 2461360.13 | 15 November 2026 |
| | | 1419.95 | 2461730.45 | 20 November 2027 |
| | | 1742.53 | 2462053.03 | 08 October 2028 |
| HD 7449 b | 5806 | 91.8 | 2460402.30 | 01 April 2024 |
| | | 1086.7 | 2461397.20 | 22 December 2026 |
| | | 1366.8 | 2461677.30 | 28 September 2027 |
| upsilon And d | 7513 | 230.89 | 2460541.39 | 18 August 2024 |
| | | 1084.9 | 2461395.40 | 20 December 2026 |
| | | 1509.02 | 2461819.52 | 18 February 2028 |
| HD 10180 h | 7599 | 34.46 | 2460344.96 | 04 February 2024 |
| | | 1143.21 | 2461453.71 | 17 February 2027 |
| HD 10647 b | 7978 | 268.16 | 2460578.66 | 25 September 2024 |
| | | 727.68 | 2461038.18 | 28 December 2025 |
| | | 1271.16 | 2461581.66 | 25 June 2027 |
| | | 1730.68 | 2462041.18 | 26 September 2028 |
| HD 13931 b | 10626 | 1356.03 | 2461666.53 | 18 September 2027 |
| epsilon Eri b | 16537 | 1105.28 | 2461415.78 | 10 January 2027 |
| HD 24040 b | 17960 | 1578.48 | 2461888.98 | 27 April 2028 |
| HD 30562 b | 22336 | 201.77 | 2460512.27 | 20 July 2024 |
| | | 282.34 | 2460592.84 | 09 October 2024 |
| | | 1358.77 | 2461669.27 | 20 September 2027 |
| | | 1439.34 | 2461749.84 | 10 December 2027 |
| GJ 179 b | 22627 | 801.6 | 2461112.10 | 12 March 2026 |
| | | 1804.45 | 2462114.95 | 09 December 2028 |
| HD 33636 b | 24205 | 761.49 | 2461071.99 | 31 January 2026 |
| | | 1571.61 | 2461882.11 | 20 April 2028 |
| HD 39091 b | 26394 | 446.51 | 2460757.01 | 22 March 2025 |
| HD 38529 c | 27253 | 481.74 | 2460792.24 | 26 April 2025 |
| | | 1396.67 | 2461707.17 | 28 October 2027 |
| 7 CMa b | 31592 | 135.7 | 2460446.20 | 15 May 2024 |
| | | 531.47 | 2460841.97 | 15 June 2025 |
| | | 898.7 | 2461209.20 | 17 June 2026 |
| | | 1294.47 | 2461604.97 | 18 July 2027 |
| | | 1661.7 | 2461972.20 | 19 July 2028 |
| HD 50499 b | 32970 | 388.84 | 2460699.34 | 23 January 2025 |
| | | 1232.0 | 2461542.50 | 17 May 2027 |
| HD 50554 b | 33212 | 118.8 | 2460429.30 | 28 April 2024 |
| | | 681.21 | 2460991.71 | 12 November 2025 |
| | | 1342.8 | 2461653.30 | 04 September 2027 |
| beta Gem b | 37826 | 115.02 | 2460425.52 | 25 April 2024 |
| | | 409.13 | 2460719.63 | 13 February 2025 |
| | | 704.66 | 2461015.16 | 05 December 2025 |
| | | 998.77 | 2461309.27 | 25 September 2026 |
| | | 1294.3 | 2461604.80 | 18 July 2027 |
| | | 1588.41 | 2461898.91 | 07 May 2028 |

Table 1. Cont.

| star name | HIP | mission day | Julian day | date |
|---|---|---|---|---|
| HD 70642 b | 40952 | 192.21 | 2460502.71 | 11 July 2024 |
|  |  | 1245.13 | 2461555.63 | 30 May 2027 |
| HD 72659 b | 42030 | 1780.78 | 2462091.28 | 15 November 2028 |
| 55 Cnc d | 43587 |  |  |  |
| GJ 328 b | 43790 | 1626.72 | 2461937.22 | 14 June 2028 |
| HD 79498 b | 45406 | 715.04 | 2461025.54 | 16 December 2025 |
|  |  | 1174.37 | 2461484.87 | 20 March 2027 |
| HD 87883 b | 49699 | 1476.61 | 2461787.11 | 16 January 2028 |
| HD 89307 b | 50473 | 680.63 | 2460991.13 | 11 November 2025 |
|  |  | 1739.6 | 2462050.10 | 05 October 2028 |
| 47 UMa b | 53721 | 303.64 | 2460614.14 | 30 October 2024 |
|  |  | 852.27 | 2461162.77 | 02 May 2026 |
|  |  | 1381.64 | 2461692.14 | 13 October 2027 |
| 47 UMa c | 53721 | 1000.37 | 2461310.87 | 27 September 2026 |
| HD 106252 b | 59610 | 504.28 | 2460814.78 | 19 May 2025 |
|  |  | 844.47 | 2461154.97 | 24 April 2026 |
| HD 117207 b | 65808 | 290.88 | 2460601.38 | 17 October 2024 |
|  |  | 1362.23 | 2461672.73 | 24 September 2027 |
| HD 128311 c | 71395 | 327.27 | 2460637.77 | 23 November 2024 |
|  |  | 724.45 | 2461034.95 | 25 December 2025 |
|  |  | 1251.07 | 2461561.57 | 05 June 2027 |
|  |  | 1648.25 | 2461958.75 | 06 July 2028 |
| HD 134987 c | 74500 | 626.59 | 2460937.09 | 18 September 2025 |
| kappa CrB b | 77655 | 386.62 | 2460697.12 | 21 January 2025 |
|  |  | 1139.06 | 2461449.56 | 13 February 2027 |
|  |  | 1686.62 | 2461997.12 | 13 August 2028 |
| HD 142022 b | 79242 | 284.5 | 2460595.00 | 11 October 2024 |
|  |  | 1380.68 | 2461691.18 | 12 October 2027 |
| 14 Her b | 79248 | 597.9 | 2460908.40 | 20 August 2025 |
|  |  | 1654.23 | 2461964.73 | 12 July 2028 |
| HD 147513 b | 80337 | 218.07 | 2460528.57 | 06 August 2024 |
|  |  | 397.53 | 2460708.03 | 01 February 2025 |
|  |  | 746.47 | 2461056.97 | 16 January 2026 |
|  |  | 925.93 | 2461236.43 | 14 July 2026 |
|  |  | 1274.87 | 2461585.37 | 28 June 2027 |
|  |  | 1454.33 | 2461764.83 | 25 December 2027 |
|  |  | 1803.27 | 2462113.77 | 08 December 2028 |
| GJ 649 b | 83043 | 67.74 | 2460378.24 | 08 March 2024 |
|  |  | 350.24 | 2460660.74 | 16 December 2024 |
|  |  | 666.04 | 2460976.54 | 28 October 2025 |
|  |  | 948.54 | 2461259.04 | 06 August 2026 |
|  |  | 1264.34 | 2461574.84 | 18 June 2027 |
|  |  | 1546.84 | 2461857.34 | 26 March 2028 |
| HD 154345 b | 83389 | 198.04 | 2460508.54 | 17 July 2024 |
| GJ 676 A b | 85647 | 239.3 | 2460549.80 | 27 August 2024 |
|  |  | 552.84 | 2460863.34 | 06 July 2025 |
|  |  | 1296.1 | 2461606.60 | 20 July 2027 |
|  |  | 1609.64 | 2461920.14 | 28 May 2028 |

Table 1. Cont.

| star name | HIP | mission day | Julian day | date |
|---|---|---|---|---|
| mu Ara c | 86796 | 490.33 | 2460800.83 | 05 May 2025 |
| mu Ara b | 86796 | 45.65 | 2460356.15 | 15 February 2024 |
| | | 387.04 | 2460697.54 | 22 January 2025 |
| | | 688.9 | 2460999.40 | 19 November 2025 |
| | | 1030.29 | 2461340.79 | 27 October 2026 |
| | | 1332.15 | 2461642.65 | 25 August 2027 |
| | | 1673.54 | 2461984.04 | 31 July 2028 |
| HD 169830 c | 90485 | 384.89 | 2460695.39 | 19 January 2025 |
| | | 1020.81 | 2461331.31 | 17 October 2026 |
| HD 181433 d | 95467 | 588.99 | 2460899.49 | 11 August 2025 |
| 16 Cyg B b | 96901 | 574.04 | 2460884.54 | 28 July 2025 |
| | | 656.5 | 2460967.00 | 18 October 2025 |
| | | 1372.54 | 2461683.04 | 04 October 2027 |
| | | 1455.0 | 2461765.50 | 25 December 2027 |
| HD 187123 c | 97336 | 463.4 | 2460773.90 | 08 April 2025 |
| | | 1823.28 | 2462133.78 | 28 December 2028 |
| HD 190360 b | 98767 | 330.16 | 2460640.66 | 26 November 2024 |
| HD 192310 c | 99825 | 152.67 | 2460463.17 | 01 June 2024 |
| | | 315.87 | 2460626.37 | 11 November 2024 |
| | | 678.47 | 2460988.97 | 09 November 2025 |
| | | 841.67 | 2461152.17 | 21 April 2026 |
| | | 1204.27 | 2461514.77 | 19 April 2027 |
| | | 1367.47 | 2461677.97 | 29 September 2027 |
| | | 1730.07 | 2462040.57 | 26 September 2028 |
| HD 196885 b | 101966 | 131.07 | 2460441.57 | 11 May 2024 |
| | | 412.54 | 2460723.04 | 16 February 2025 |
| | | 1464.07 | 2461774.57 | 04 January 2028 |
| | | 1745.54 | 2462056.04 | 11 October 2028 |
| HD 204313 d | 106006 | 863.27 | 2461173.77 | 13 May 2026 |
| HD 204941 b | 106353 | 798.23 | 2461108.73 | 09 March 2026 |
| | | 1355.88 | 2461666.38 | 17 September 2027 |
| GJ 832 b | 106440 | 85.56 | 2460396.06 | 26 March 2024 |
| | | 1577.08 | 2461887.58 | 26 April 2028 |
| GJ 849 b | 109388 | 611.61 | 2460922.11 | 03 September 2025 |
| | | 1556.79 | 2461867.29 | 05 April 2028 |
| HD 216437 b | 113137 | 50.04 | 2460360.54 | 20 February 2024 |
| | | 978.82 | 2461289.32 | 05 September 2026 |
| | | 1403.04 | 2461713.54 | 04 November 2027 |
| HD 217107 c | 113421 | 898.64 | 2461209.14 | 17 June 2026 |
| HD 220773 b | 115697 | 717.43 | 2461027.93 | 18 December 2025 |
| | | 1439.43 | 2461749.93 | 10 December 2027 |
| HD 222155 b | 116616 | 357.83 | 2460668.33 | 23 December 2024 |
| gamma Cep b | 116727 | 356.11 | 2460666.61 | 22 December 2024 |
| | | 861.67 | 2461172.17 | 11 May 2026 |
| | | 1261.69 | 2461572.19 | 15 June 2027 |
| | | 1767.24 | 2462077.74 | 02 November 2028 |

## 2.2 SIGNAL-TO-NOISE RATIO, S/N

Two sub-theories are required to estimate the value of S/N required by Equation (1) to achieve the desired mass accuracy. In §2.2.1, we derive the relationship between S/N and detector counts—signal counts $\mathcal{S}$ and noise counts $\mathcal{N}$—for the roll-deconvolution protocol, which we adopt for this study. In §2.2.2, we empirically estimate the relationship between astrometric error $\sigma_s$ and S/N. Later, in §3.3, we will show how to compute the exposure time $\tau$. At that point, we will have a three-step bridge between Equation (1) and the required value of $\tau$, which we will be able to compute for any RV planet, on any mission day, for any inclination $i$.

### 2.2.1 ROLL DECONVOLUTION, S/N, $\mathcal{S}$, and $\mathcal{N}$

"Roll deconvolution" is a scheme to reduce the effect of speckles, which—assuming the optics are stable—do not move when the telescope is rolled around the optic axis by an angle $\phi$ in the time between the two otherwise equal exposures (e.g. Müller & Weigelt 1987). Assume $(x, y)$ coordinates fixed on the focal-plane detector, with origin $(0, 0)$ located on the optic axis of the telescope at the position of the host star. Assume the two exposures produce two data arrays—images $I_1$ and $I_2$—each with half the total exposure time, $\tau$. The position in $I_2$ of a point located at position $(x_1, y_1)$ in $I_1$ is:

$$(x_2, y_2) = (x_1 \cos\varphi - y_1 \sin\varphi, x_1 \sin\varphi + y_1 \cos\varphi). \tag{13}$$

Data analysis for the roll-convolution protocol involves the following the following steps.

Step 1. Subtract $I_2$ from $I_1$, creating the difference image $I_3$. The purpose of this step is to remove the speckles, which are not moved by the roll, leaving only speckle noise.

If we hypothesize that a planet is located at $(x_1, y_1)$ in $I_1$, we expect to find the value $+\mathcal{S}/2$ signal photons to be collected in the optimal virtual photometric aperture (VPA) of $\Psi^{-1} = 13$ Nyquist-sampling pixels centered at $(x_1, y_1)$ in $I_3$ (Burrows, 2003, Burrows, Brown, & Sabatke, 2006).

If the VPA is centered at $(x_2, y_2)$ in $I_3$, we expect to find the value $-\mathcal{S}/2$. In the field of $I_3$, well away from $(x_1, y_1)$ and $(x_2, y_2)$, we expect to find noise of zero mean and standard deviation $\sqrt{\mathcal{N}}$, where we expect $\mathcal{N}/2$ noise photons to be collected in the field for each of $I_1$ and $I_2$.

Step2. Perform the reverse transformation (rotation) on $I_3$, creating image $I_4$.

Step 3. Subtract image $I_4$ from $I_3$, creating the final image, $I_5$. We now find $\mathcal{S}$ signal photons in the VPA centered at $(x_1, y_1)$ in $I_5$. (We ignore the two negative, ghost images on either side.) In the field, the standard deviation of the noise in the VPA is now $\sqrt{2\mathcal{N}}$. Therefore, the S/N for roll deconvolution is

$$\text{S/N} = \frac{\mathcal{S}}{\sqrt{\mathcal{S} + 2\mathcal{N}}}. \tag{14}$$

### 2.2.2 EMPIRICAL DETERMINATION OF S/N REQUIRED BY $\sigma_s$

A single roll-deconvolved image (e.g. $I_5$) of a known RV planet yields estimates of three quantities: (1) the brightness of the planet relative to the star (expressed in stellar magnitudes as $\Delta mag$) and (2–3) the planet's two-dimensional position, $(x, y)$, also relative to the star. With the origin of coordinates positioned on the star, the apparent separation of the planet is $s \equiv \sqrt{x^2 + y^2}$. In this section, we use Monte Carlo experiments to estimate $\sigma_s$, the standard deviation or uncertainty in $s$, as a function of S/N—or equivalently S/N as a function of $\sigma_s$, as needed. These experiments use least-squares fits to estimate the position and brightness of the planet in noisy, simulated images, which serve as analogs to the roll-deconvolved images expected in the missions.

In these experiments, we use the two control parameters S/N and the count ratio

$$cr \equiv \mathcal{S}/\mathcal{N}, \tag{15}$$

which are equivalent to $\mathcal{S}$ and $\mathcal{N}$ as control parameters, because of Equation (14).

The Monte Carlo experiments to determine the relationship between $\sigma_s$, S/N, and $cr$ involve eight steps. (1) We define a 15-by-15 detector array with critically sampling pixels of width $\lambda/2D'$ radians, where $D'$ is the diameter of the *exit* pupil. (2) We produce a *pixel-integrated* point-spread function (piPSF) centered on $(x, y) = (0, 0)$, and then, using Whittaker–Shannon interpolation, we translate piPSF to a random location $(x_{\text{true}}, y_{\text{true}})$ on the detector in the range $-4 < x_{\text{true}}, y_{\text{true}} < 4$ pixels. (3) For choices of control parameters, S/N and $cr$, we compute $\mathcal{S}$ and $\mathcal{N}$ using Equations (14–15). (4) We multiply piPSF by $\mathcal{S}$ and then replace each pixel's resulting value of piPSF $\times \mathcal{S}$ by a number drawn from a Poisson random deviate with a mean equal to that value divided by $vpae = 0.85$, which is the efficiency of the VPA—i.e. the sum of piPSF over the 13 pixels. This correction ensures that, on average, $\mathcal{S}$ photons will be collected in the VPA, as intended. (5) We introduce background noise by adding to each pixel in the detector a value drawn from the random deviate for the Skellam distribution, both parameters of which have been set equal to $\mathcal{N}/13$, which is the mean value of background noise counts per pixel. The result emulates the noise in the difference of two noisy images. (6) We perform a least squares fit to estimate the values of $\mathcal{S}, \mathcal{N}$—which yields S/N—and $(x_{\text{fit}}, y_{\text{fit}})$. (7) We estimate the astrometric error as the distance of $(x_{\text{fit}}, y_{\text{fit}})$ from $(x_{\text{true}}, y_{\text{true}})$:

$$s_{\text{error}} \equiv \sqrt{(x_{\text{fit}} - x_{\text{true}})^2 + (y_{\text{fit}} - y_{\text{true}})^2} \tag{16}$$

(8) We perform steps 1–7 many times with fresh realizations of the noise and true image location each time, and from the results, estimate $\sigma_s$ as the root-mean-square value of $s_{\text{error}}$.

The Monte Carlo trials produce two immediate findings. First, we find that the results of this experiment are independent of $cr$ in the range $10^{-1}$–$10^{-4}$, and we therefore adopt the single value $cr = 10^{-2}$ for ongoing computations. Second, in 10,000 Monte Carlo trials for various choices of S/N < 8, we find that the outcomes of the least squares fit depend—no surprise—on the quality of the initial guess of the planetary location. This initial guess is the starting point of the search for a global minimum of the objective function. The only reasonable option is to choose the position of the maximum value in the data array as that initial guess.

As S/N decreases, a growing fraction of the peak values will be noise fluctuations remote from and unrelated to the planet. This effect increases the probability that the least squares fit will not find the planetary image but will stall on a local minimum of pure noise. When this happens, the result is an outlier value of $s_{error}$. We want to understand these outliers—and hopefully minimize or prevent them, because in real data they might be indistinguishable from a faint planetary signal.

We document the outlier issue as follows. For each of 18 values of $S/N_{img}$ in the range 2–100, we produce 10,000 values of $s_{error}$ by steps 1–7 above. Now, if we knew the value of $\sigma_s$ —which we do not, and in fact we are trying to determine it—we would expect $(s_{error} / \sigma_s)^2$ to be distributed as chi squared with two degrees of freedom. For $S/N < 8$, however, a considerable fraction of Monte Carlo trials produce outliers with probabilities less than $10^{-4}$ of being drawn from the expected chi-squared distribution. On this basis, we set the cutoff $(s_{error} / \sigma_s)^2 > 18$ (i.e. false-negative probability of $10^{-4}$) to remove outliers. When planning exposure times, we require $S/N \geq 8$.

We estimate $\sigma_s$ and purge the outliers according to the following self-consistent process. We define an objective function equal to the Kolmogorov-Smirnov test statistic for the case where (a) the data set is the sample of 10,000 values of $s_{error}^2$ with each value divided by $\sigma_s$—here treated here as a parameter—and with values $(s_{error} / \sigma_s)^2 > 18$ removed, and (b) the reference distribution is the chi-square distribution with two degrees of freedom. We then perform a least squares fit to find the value of $\sigma_s$ that minimizes the objective function. The results are shown in Figs. 1–2.

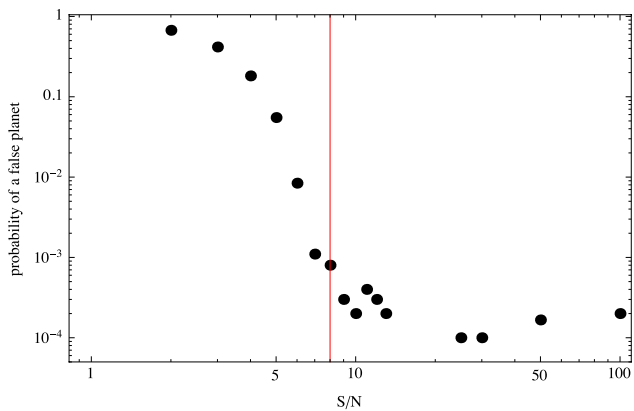

**Figure 1.** Fraction of outliers versus S/N in 10, 000 Monte Carlo trials at each point. Red line: adopted minimum value S/N = 8, which is the cutoff to avoid outliers.

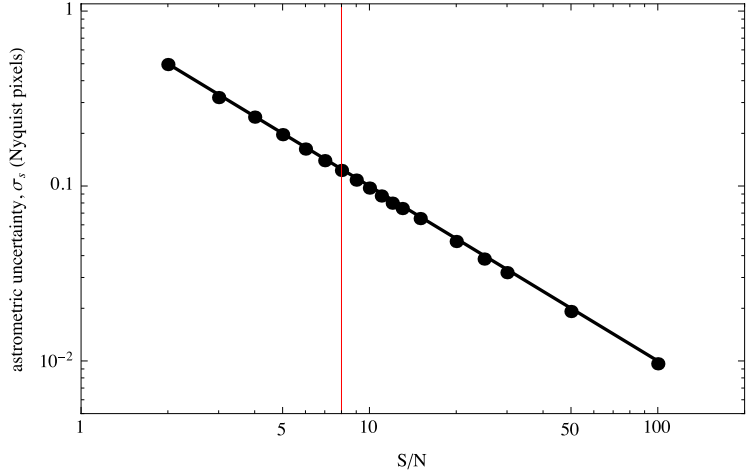

**Figure 2.** Astrometric uncertainty $\sigma_s$ in Nyquist pixels as a function of S/N. Each point is the result of a least-squares fit to minimize the Kolmogorov-Smirnov test statistic computed for a data set of 10, 000 values of $s_{error}$ minus outliers. Black line: the function $(S/N)^{-1}$. Red line: minimum value $S/N = 8$, which is the cutoff to avoid outliers.

For $\Delta/\sigma_s > 8$, where $\Delta$ is the width of a Nyquist pixel in arcsec:

$$\Delta = \frac{\lambda}{2D' \, 4.848 \times 10^{-6}}, \tag{17}$$

the requirement is

$$S/N = \frac{\Delta}{\sigma_s} = \frac{s \, \Delta}{x_0^2 + y_0^2 - s^2} \left(\frac{\sigma_{m_p}}{m_p}\right)^{-1} \tag{18}$$

and for $\Delta/\sigma_s \leq 8$, we have

$$S/N = 8. \tag{19}$$

In §3.3, we describe the process of estimating the exposure time $\tau$ needed to achieve the required accuracy in the planetary mass. The calibration in Figure 2 and Equation 18–19) play essential roles in that process. For each planet in its own unique way, depending on its orbital parameters and orbital phase, the desired mass uncertainty begets a maximum value of the astrometric uncertainty $\sigma_s$, which begets the minimum value of S/N, which begets the required exposure time, $\tau$—for any set of design parameters, any Julian day, and any random value of inclination $i$.

# 3. MISE EN PLACE

In §3.1, we summarize the design parameters that define our "baseline" performance estimates of science outcomes. In §3.2, we describe our input catalog of known RV planets. In §3.3, we discuss the calculation of exposure time $\tau$. In §4, we introduce the cube of intermediate information that realizes a mission. In §5, we introduce the concept of the design reference mission (DRM). In §6 we show the DRM results for the four missions under study.

## 3.1 BASELINE PARAMETERS AND ASSUMPTIONS

Table 2 documents the baseline assumptions and parameters for the four missions under study, and indicates whether we regard them as instrumental (defining the observing system), astronomical (defining the sky), or science-operational parameters (indicating how we would expect to use the observatory to investigate known RV planets. Our instrumental parameters are consistent with the current engineering studies for those missions (Shaklan, private communication).

As in Brown (2015), we adopt the operational wavelength $\lambda = 760$ nm (*I* band), and resolving powers $\mathcal{R} = 5$ (imaging) and $\mathcal{R} = 70$ (spectroscopy). These choices reflect our view of the priority of the diagnostic features of methane and molecular oxygen that occur in the range 700–820 nm. The spectroscopic resolving power of 70 is reasonably matched to the expected widths of those features (Des Marais et al. 2002).

The diameter of the *exit* pupil, $D´$, which contributes to the-end to-end efficiency and governs the image size, is less than $D$ for *EXO-C* due to the Lyot mask. *EXO-S* and *WFIRST-S* have no Lyot mask, so $D´ = D$. Even though the *WFIRST-C* includes a Lyot mask, the correct choice is $D´ = D$, for that mission also, due to the cancelling effects of the inner and outer radii of the exit pupil. That is, the large *inner* radius of the Lyot mask in *WFIRST-C* removes low spatial frequencies in favor of high frequencies, which tends to reduce image size, while the smaller *outer* radius, $D´ < D$, tends to increase it. As a result, in this case, we have learned that the central portion of the point-spread function for *WFIRST-C* does not differ by more than a few percent from that of an idealized 2.4 m telescope with $D´ = D$ (Krist, private communication). Therefore, we use $D´ = D = 2.4$ m for *WFIRST-C* to ensure the correct size of a Nyquist pixel and virtual photometric aperture VPA, and we introduce a "Lyot mask correction" ($LMC = 0.08$) to ensure we include Lyot mask losses—only for the case of *WFIRST-C*.

We assume that the pixel width ($\Delta$) critically samples the planetary image on the detector according to the Nyquist criterion, $\Delta = \lambda / 2D´$.

For *EXO-S* and *EXO-C*, we assume a CCD detector with dark noise $\upsilon = 0.00055$, read noise $\sigma = 2.8$, read-out cadence $t_r = 2000$ sec, and quantum efficiency $\varepsilon = 0.8$, which are the values used in current NASA studies. For *WFIRST*, we assume a photon-counting detector with $\upsilon = 0$, $\sigma = 0$, and $\varepsilon = 0.8$.

We assume sufficiently stable optics to support speckle disambiguation and noise reduction by roll deconvolution, as described in §2.2.1.

Scattered starlight is suppressed to a level specified by $\zeta$, defined as the ratio of the intensity of scattered starlight in the detection zone to the central intensity of the theoretical Airy image of the star.

**Table 2.** Baseline parameters

| # | Parameter | Symbol | Value | Units | Comment | Type |
|---|---|---|---|---|---|---|
| 1 | wavelength | $\lambda$ | 760 | nm | | science operational |
| 2 | spectral resolving power imaging | $\mathcal{R}_{img}$ | 5 | | | science operational |
| 3 | spectral resolving power spectroscopy | $\mathcal{R}_{spc}$ | 70 | | | science operational |
| 4 | entrance aperture | $D$ | 1.1 *EXO-S* <br> 1.4 *EXO-C* <br> 2.4 *WFIRST-C, -S* | m | | instrumental |
| 5 | exit aperture | $D'$ | 1.1 *EXO-S* <br> 0.85 *EXO-C* <br> 2.4 *WFIRST-C, -S* | m | | Instrumental |
| 6 | pixel width | $\Delta$ | 0.071 *EXO-S* <br> 0.092 *EXO-C* <br> 0.033 *WFIRST-C, -S* | arcsec | Nyquist: $\lambda / 2 D'$ | dependent parameter |
| 7 | dark noise | $\upsilon$ | 0.00055 *EXO-S* <br> 0.00055 *EXO-C* <br> 0. *WFIRST-C, -S* | | | instrumental |
| 8 | read noise | $\sigma$ | 2.8 *EXO-S* <br> 2.8 *EXO-C* <br> 0. *WFIRST-C, -S* | | | instrumental |
| 9 | readout cadence | $t_r$ | 2000 *EXO-S* <br> 2000 *EXO-C* <br> N/A *WFIRST-C, -S* | sec | | science operational |
| 10 | inner working angle | IWA | 0.100 *EXO-S* <br> 0.224 *EXO-C* <br> 0.100 *WFIRST-S* <br> 0.196 *WFIRST-C* | arcsec | | instrumental |
| 11 | scattered starlight | $\zeta$ | 5 x 10$^{-10}$ *EXO-S* <br> 10$^{-9}$ *EXO-C* <br> 10$^{-10}$ *WFIRST-S* <br> 10$^{-9}$ *WFIRST-C* | Airy intensity of stellar image center | | instrumental |
| 12 | Lyot mask correction | LMC | 1.0 *EXO-S* <br> 1.0 *EXO-C* <br> 1.0 *WFIRST-S* <br> 0.08 *WFIRST-C* | | corrects Lyot mask throughput | instrumental |
| 13 | detector QE | $\varepsilon$ | 0.8 | | | instrumental |
| 14 | refl/trans efficiency | reflTrans | 0.45 *EXO-S* <br> 0.4 *EXO-C* <br> 0.25 *WFIRST-S, -C* | | | instrumental |
| 15 | planet radius | $R$ | 1 | $R_{Jupiter}$ | | astronomical |
| 16 | geometric albedo | $p$ | 0.5 | | | astronomical |
| 17 | phase function | $\Phi$ | Lambertian | | | astronomical |
| 18 | mission start | $t_0$ | 2 460 310.5 | JD | 1/1/2024 | science operational |

**Table 2. Baseline parameters (cont.)**

| # | Parameter | Symbol | Value | Units | Comment | Type |
|---|---|---|---|---|---|---|
| 19 | solar avoidance | $\gamma_1$ | 30° *EXO-S*<br>45° *EXO-C*<br>54° *WFIRST-C*<br>40° *WFIRST-S* | deg | | instrumental |
| 20 | antisolar avoidance | $\gamma_2$ | 83° *EXO-S*<br>180° *EXO-C*<br>83° *WFIRST-S*<br>126° *WFIRST-C* | deg | | instrumental |
| 21 | setup time | $t_{setup}$ | 8 *EXO-S*<br>0.5 *EXO-C*<br>8 *WFIRST-S*<br>0.5 *WFIRST-C* | days | per observation | science operational |
| 22 | mission duration | $t_{mission}$ | 730 *EXO-S, -C*<br>730 *WFIRST-S*<br>1095 *WFIRST-C* | days | | science operational |
| 23 | zodiacal light | $Z$ | 7 | zodis | 23$^{rd}$ magnitude arcsec$^{-2}$ | astronomical |
| 24 | limiting delta magnitude | $\Delta mag_0$ | 22.5 | magnitudes | | science operational |
| 25 | *S/N*, imaging | $S/N_{img}$ | 8 | | | science operational |
| 26 | *S/N*, spectroscopy | $S/N_{spc}$ | 10 | | | science operational |
| 27 | mass accuracy | $\sigma_{mp}/m_p$ | 0.10 | | | science operational |
| 28 | sharpness | $\Psi$ | 0.08 | | | instrumental |
| 29 | zero-magnitude flux | $F_0$ | 4885 | photons cm$^{-2}$ nm$^{-1}$ sec$^{-1}$ | *I* band, 760 nm | science operational |

**Note:** For the coronagraphs, *IWA* = 3 $\lambda$ / *D* for *WFIRST-C*, and *IWA* = 2 $\lambda$ / *D* for *EXO-C*.

## 3.2 INPUT CATALOG

Our input catalog of known radial-velocity planets for each mission was selected from the list at www.exoplanets.org using the criterion that the raw angular separation $s_{raw} \equiv a(1+e)/d > IWA$, where $a$ is the semimajor axis, $e$ is the orbital eccentricity, and $d$ is the stellar distance. $s_{raw}$ is the maximum possible apparent separation for a planet with parameters $a$, $e$, and $d$. It does not take into account the orientation of the orbit, particularly $\omega_p$, the argument of periastron of the planet. If $e \neq 0$, $\omega_p$ generally reduces the maximum *achievable* separation ($s_{max}$) to a smaller value, $s_{max} < s_{raw}$.

Fig. 3 shows the stars in input catalogs and the restrictions of the adopted values of $s_1$ for *WFIRST-C*, *WFIRST-S*, *EXO–C*, and *EXO–S*.

Table 3 shows the union input catalog of RV planets and host stars, including a working selection of data items for each exoplanet.

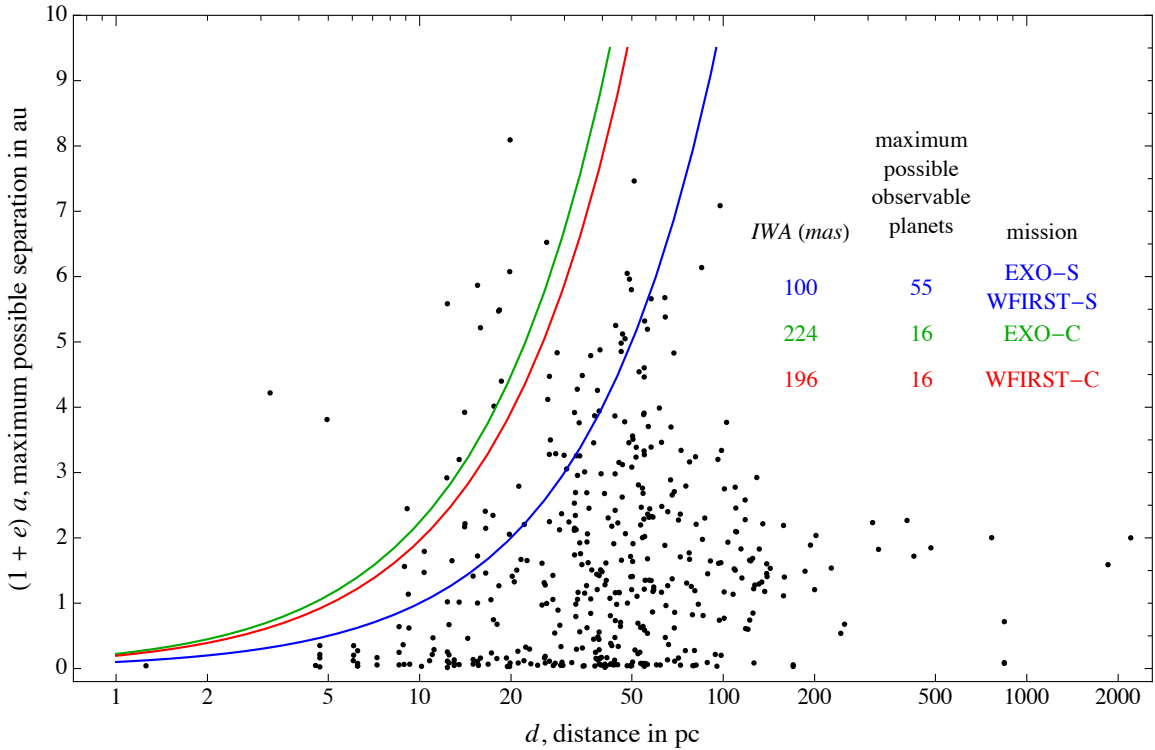

**Figure 3** Distance ($d$) and value of raw separation $a(1+e)$ for the 436 known radial-velocity planets in the data base at www.exoplanets.org on March 1, 2014. The curves parse the planets into two groups, those to the right, which are totally obscured for that value of $s_1$ and therefore never observable, and those to the left, which *may* be observable at some time during the mission, depending on the orbital inclination, $i$. Blue: *EXO-S*, green: *WFIRST-C*, red: *EXO-C*.

**Table 3. Union input catalogs of RV planets and host stars.**

| # | coronagraphs | star shades | planet name | HIP | V | B–V | I | d (pc) | mass (solar) | RA (hours) | Dec (°) | ecliptic longitude (°) | ecliptic latitude (°) | raw separation (arcsec) | a (au) | period (d) | e | ω star (°) | time of periapsis (JD) | m sin i |
|---|---|---|---|---|---|---|---|---|---|---|---|---|---|---|---|---|---|---|---|---|
| 1 | | ✓ | GJ 317 b | … | 12 | 1.52 | 9.63 | 9.1743 | 0.24 | 8.7 | -23.5 | 140.8 | -40.1 | 0.124 | 0.95 | 692.9 | 0.19 | 344 | 2451639 | 1.18 |
| 2 | | ✓ | HD 7449 b | 5806 | 7.5 | 0.58 | 6.88 | 38.521 | 1.05 | 1.2 | -5.0 | 15.2 | -12.0 | 0.1105 | 2.34 | 1275 | 0.82 | 339 | 2455298 | 1.31 |
| 3 | ✓ | ✓ | upsilon And d | 7513 | 4.1 | 0.54 | 3.51 | 13.492 | 1.31 | 1.6 | 41.4 | 38.6 | 29.0 | 0.2372 | 2.52 | 1278 | 0.27 | 270 | 2453937 | 4.12 |
| 4 | | ✓ | HD 10180 h | 7599 | 7.33 | 0.63 | 6.65 | 39.017 | 1.06 | 1.6 | -60.5 | 340.4 | -61.6 | 0.101 | 3.42 | 2248 | 0.15 | 184 | 2453619 | 0.21 |
| 5 | | ✓ | HD 10647 b | 7978 | 5.52 | 0.55 | 4.92 | 17.434 | 1.1 | 1.7 | -53.7 | 350.8 | -57.3 | 0.1345 | 2.02 | 1003 | 0.16 | 336 | 2450960 | 0.93 |
| 6 | | ✓ | HD 13931 b | 10626 | 7.61 | 0.64 | 6.92 | 44.228 | 1.02 | 2.3 | 43.8 | 47.3 | 28.2 | 0.1188 | 5.15 | 4218 | 0.02 | 290 | 2454494 | 1.88 |
| 7 | ✓ | ✓ | epsilon Eri b | 16537 | 3.72 | 0.88 | 2.79 | 3.2161 | 0.82 | 3.5 | -9.5 | 48.2 | -27.7 | 1.312 | 3.38 | 2500 | 0.25 | 6 | 2448940 | 1.05 |
| 8 | | ✓ | HD 24040 b | 17960 | 7.5 | 0.65 | 6.8 | 46.598 | 1.18 | 3.8 | 17.5 | 59.2 | -2.6 | 0.1099 | 4.92 | 3668 | 0.04 | 154 | 2454308 | 4.02 |
| 9 | | ✓ | HD 30562 b | 22336 | 5.77 | 0.63 | 5.09 | 26.42 | 1.28 | 4.8 | -5.7 | 69.8 | -27.9 | 0.1559 | 2.34 | 1157 | 0.76 | 81 | 2450131 | 1.33 |
| 10 | ✓ | ✓ | GJ 179 b | 22627 | 11.96 | 1.6 | 9.15 | 12.288 | 0.36 | 4.9 | 6.5 | 72.4 | -15.9 | 0.2376 | 2.41 | 2288 | 0.21 | 153 | 2455140 | 0.82 |
| 11 | | ✓ | HD 33636 b | 24205 | 7 | 0.59 | 6.36 | 28.369 | 1.02 | 5.2 | 4.4 | 77.3 | -18.5 | 0.1704 | 3.27 | 2128 | 0.48 | 340 | 2451205 | 9.27 |
| 12 | ✓ | ✓ | HD 39091 b | 26394 | 5.65 | 0.6 | 5 | 18.315 | 1.07 | 5.6 | -80.5 | 273.9 | -76.0 | 0.2998 | 3.35 | 2151 | 0.64 | 330 | 2447820 | 10.09 |
| 13 | | ✓ | HD 38529 c | 27253 | 5.95 | 0.77 | 5.13 | 39.277 | 1.34 | 5.8 | 1.2 | 86.4 | -22.2 | 0.1242 | 3.6 | 2146 | 0.36 | 17.9 | 2452255 | 12.26 |
| 14 | | ✓ | 7 CMa b | 31592 | 3.95 | 1.06 | 2.83 | 19.751 | 1.34 | 6.6 | -19.3 | 101.7 | -42.3 | 0.104 | 1.8 | 763 | 0.14 | 12 | 2455520 | 2.43 |
| 15 | | ✓ | HD 50499 b | 32970 | 7.21 | 0.61 | 6.55 | 46.126 | 1.28 | 6.9 | -33.9 | 109.8 | -56.5 | 0.1052 | 3.87 | 2458 | 0.25 | 259 | 2451220 | 1.74 |
| 16 | | ✓ | HD 50554 b | 33212 | 6.84 | 0.58 | 6.21 | 29.913 | 1.03 | 6.9 | 24.2 | 102.5 | 1.4 | 0.1091 | 2.26 | 1224 | 0.44 | 7.4 | 2450646 | 4.40 |
| 17 | | ✓ | beta Gem b | 37826 | 1.15 | 1 | 0.1 | 10.358 | 2.08 | 7.8 | 28.0 | 113.2 | 6.7 | 0.1731 | 1.76 | 589.6 | 0.02 | 355 | 2447739 | 2.76 |
| 18 | | ✓ | HD 70642 b | 40952 | 7.17 | 0.69 | 6.43 | 28.066 | 1 | 8.4 | -39.7 | 144.2 | -56.7 | 0.1172 | 3.18 | 2068 | 0.03 | 205 | 2451350 | 1.91 |
| 19 | | ✓ | HD 72659 b | 42030 | 7.46 | 0.61 | 6.8 | 49.826 | 1.07 | 8.6 | -1.6 | 131.4 | -19.6 | 0.1164 | 4.75 | 3658 | 0.22 | 261 | 2455351 | 3.17 |
| 20 | ✓ | ✓ | 55 Cnc d | 43587 | 5.96 | 0.87 | 5.04 | 12.341 | 0.91 | 8.9 | 28.3 | 127.7 | 10.4 | 0.4525 | 5.47 | 4909 | 0.02 | 254 | 2453490 | 3.54 |
| 21 | ✓ | ✓ | GJ 328 b | 43790 | 9.98 | 1.32 | 8.41 | 19.8 | 0.69 | 8.9 | 1.5 | 135.8 | -15.2 | 0.3068 | 4.43 | 4100 | 0.37 | 290 | 2454500 | 2.30 |
| 22 | | ✓ | HD 79498 b | 45406 | 8.02 | 0.71 | 7.26 | 46.104 | 1.06 | 9.3 | 23.4 | 134.1 | 7.1 | 0.108 | 3.13 | 1966 | 0.59 | 221 | 2453210 | 1.35 |
| 23 | ✓ | ✓ | HD 87883 b | 49699 | 7.57 | 0.96 | 6.56 | 18.205 | 0.8 | 10.1 | 34.2 | 141.7 | 21.3 | 0.3006 | 3.58 | 2754 | 0.53 | 291 | 2451139 | 1.76 |
| 24 | | ✓ | HD 89307 b | 50473 | 7.02 | 0.59 | 6.38 | 32.363 | 0.99 | 10.3 | 12.6 | 151.9 | 1.9 | 0.1211 | 3.27 | 2166 | 0.20 | 4.9 | 2452346 | 1.79 |
| 25 | ✓ | ✓ | 47 UMa c | 53721 | 5.03 | 0.62 | 4.36 | 14.063 | 1.06 | 11.0 | 40.4 | 149.1 | 31.1 | 0.2789 | 3.57 | 2391 | 0.10 | 295 | 2452441 | 2.55 |
| 26 | | ✓ | 47 UMa b | " | " | " | " | " | " | " | " | " | " | 0.1542 | 2.1 | 1078 | 0.03 | 334 | 2451917 | 0.55 |
| 27 | | ✓ | HD 106252 b | 59610 | 7.41 | 0.64 | 6.73 | 37.707 | 1.01 | 12.2 | 10.0 | 179.1 | 10.5 | 0.1026 | 2.61 | 1531 | 0.48 | 293 | 2453397 | 6.96 |
| 28 | | ✓ | HD 117207 b | 65808 | 7.26 | 0.72 | 6.49 | 33.047 | 1.03 | 13.5 | -35.6 | 214.4 | -24.3 | 0.1294 | 3.74 | 2597 | 0.14 | 73 | 2450630 | 1.82 |
| 29 | | ✓ | HD 128311 c | 71395 | 7.48 | 0.97 | 6.46 | 16.502 | 0.83 | 14.6 | 9.7 | 213.2 | 23.7 | 0.1301 | 1.75 | 923.8 | 0.23 | 28 | 2456987 | 3.25 |
| 30 | ✓ | ✓ | HD 134987 c | 74500 | 6.47 | 0.69 | 5.73 | 26.206 | 1.05 | 15.2 | -25.3 | 232.8 | -7.1 | 0.249 | 5.83 | 5000 | 0.12 | 195 | 2451100 | 0.80 |
| 31 | | ✓ | kappa CrB b | 77655 | 4.79 | 1 | 3.74 | 30.497 | 1.58 | 15.9 | 35.7 | 222.7 | 53.9 | 0.1002 | 2.72 | 1300 | 0.13 | 83.1 | 2453899 | 1.97 |
| 32 | | ✓ | HD 142022 b | 79242 | 7.7 | 0.79 | 6.86 | 34.329 | 0.9 | 16.2 | -84.2 | 264.5 | -61.3 | 0.1307 | 2.93 | 1928 | 0.53 | 170 | 2450941 | 4.47 |
| 33 | ✓ | ✓ | 14 Her b | 79248 | 6.61 | 0.88 | 5.68 | 17.572 | 1.07 | 16.2 | 43.8 | 223.2 | 62.9 | 0.2286 | 2.93 | 1773 | 0.37 | 22.6 | 2451372 | 5.21 |
| 34 | | ✓ | HD 147513 b | 80337 | 5.37 | 0.63 | 4.7 | 12.778 | 1.07 | 16.4 | -39.2 | 250.7 | -17.3 | 0.1291 | 1.31 | 528.4 | 0.26 | 282 | 2451123 | 1.18 |
| 35 | | ✓ | GJ 649 b | 83043 | 9.7165 | 1.52 | 7.34 | 10.345 | 0.54 | 17 | 25.7 | 248.9 | 48.1 | 0.1422 | 1.13 | 598.3 | 0.30 | 352 | 2452876 | 0.33 |
| 36 | ✓ | ✓ | HD 154345 b | 83389 | 6.76 | 0.73 | 5.98 | 18.587 | 0.89 | 17.0 | 47.1 | 241.7 | 69.1 | 0.2367 | 4.21 | 3342 | 0.04 | 68.2 | 2452831 | 0.96 |
| 37 | | ✓ | GJ 676 A b | 85647 | 9.585 | 1.44 | 7.68 | 16.45 | 0.71 | 17.5 | -51.6 | 264.8 | -28.3 | 0.1463 | 1.82 | 1057 | 0.33 | 85.7 | 2455411 | 4.90 |
| 38 | ✓ | ✓ | mu Ara c | 86796 | 5.12 | 0.69 | 4.38 | 15.511 | 1.15 | 17.7 | -51.8 | 267.2 | -28.4 | 0.3782 | 5.34 | 4206 | 0.10 | 57.6 | 2452955 | 1.75 |
| 39 | | ✓ | mu Ara b | " | " | " | " | " | " | " | " | " | " | 0.1111 | 1.53 | 643.3 | 0.13 | 22 | 2452365 | 1.89 |
| 40 | | ✓ | HD 169830 c | 90485 | 5.9 | 0.52 | 5.33 | 36.603 | 1.41 | 18.5 | -29.8 | 276.1 | -6.5 | 0.1309 | 3.6 | 2102 | 0.33 | 252 | 2452516 | 4.06 |
| 41 | | ✓ | HD 181433 d | 95467 | 8.38 | 1.01 | 7.32 | 26.759 | 0.78 | 19.4 | -66.5 | 281.6 | -43.9 | 0.1671 | 3.02 | 2172 | 0.48 | 330 | 2452154 | 0.54 |
| 42 | | ✓ | 16 Cyg B b | 96901 | 6.25 | 0.66 | 5.54 | 21.213 | 0.96 | 19.7 | 50.5 | 321.2 | 69.5 | 0.1316 | 1.66 | 798.5 | 0.68 | 85.8 | 2446549 | 1.64 |
| 43 | | ✓ | HD 187123 c | 97336 | 7.83 | 0.66 | 7.12 | 48.263 | 1.04 | 19.8 | 34.4 | 309.5 | 54.3 | 0.1253 | 4.83 | 3806 | 0.25 | 243 | 2453580 | 1.94 |
| 44 | ✓ | ✓ | HD 190360 b | 98767 | 5.73 | 0.75 | 4.93 | 15.858 | 0.98 | 20.1 | 29.9 | 312.6 | 48.9 | 0.329 | 3.97 | 2915 | 0.31 | 12.9 | 2453541 | 1.54 |
| 45 | | ✓ | HD 192310 c | 99825 | 5.73 | 0.88 | 4.8 | 8.9111 | 0.8 | 20.3 | -27.0 | 300.0 | -7.0 | 0.1753 | 1.18 | 525.8 | 0.32 | 110 | 2455311 | 0.07 |
| 46 | | ✓ | HD 196885 b | 101966 | 6.39 | 0.55 | 5.83 | 33.523 | 1.23 | 20.7 | 11.2 | 315.8 | 28.6 | 0.1122 | 2.54 | 1333 | 0.48 | 78 | 2452554 | 2.94 |
| 47 | | ✓ | HD 204313 d | 106006 | 8.006 | 0.7 | 7.26 | 47.483 | 1.02 | 21.5 | -21.7 | 317.5 | -6.5 | 0.1063 | 3.94 | 2832 | 0.28 | 247 | 2446376 | 1.61 |
| 48 | | ✓ | HD 204941 b | 106353 | 8.45 | 0.88 | 7.52 | 26.954 | 0.74 | 21.5 | -21.0 | 318.7 | -6.0 | 0.1298 | 2.55 | 1733 | 0.37 | 228 | 2456015 | 0.27 |
| 49 | ✓ | ✓ | GJ 832 b | 106440 | 8.66 | 1.52 | 6.29 | 4.9537 | 0.45 | 21.6 | -49.0 | 308.6 | -32.5 | 0.7694 | 3.4 | 3416 | 0.12 | 304 | 2451211 | 0.64 |
| 50 | ✓ | ✓ | GJ 849 b | 109388 | 10.42 | 1.52 | 8.05 | 9.0959 | 0.49 | 22.2 | -4.6 | 332.7 | 6.3 | 0.2691 | 2.35 | 1882 | 0.04 | 355 | 2451488 | 0.83 |
| 51 | | ✓ | HD 216437 b | 113137 | 6.04 | 0.66 | 5.33 | 26.745 | 1.12 | 22.9 | -70.1 | 305.3 | -55.5 | 0.1226 | 2.49 | 1353 | 0.32 | 67.7 | 2450605 | 2.17 |
| 52 | ✓ | ✓ | HD 217107 c | 113421 | 6.17 | 0.74 | 5.38 | 19.857 | 1.11 | 23.0 | -2.4 | 344.9 | 3.9 | 0.4076 | 5.33 | 4270 | 0.52 | 199 | 2451106 | 2.62 |
| 53 | | ✓ | HD 220773 b | 115697 | 7.06 | 0.66 | 6.35 | 50.891 | 1.16 | 23.4 | 8.6 | 355.8 | 11.3 | 0.1467 | 4.94 | 3725 | 0.51 | 259 | 2453866 | 1.45 |
| 54 | | ✓ | HD 222155 b | 116616 | 7.12 | 0.64 | 6.43 | 49.1 | 1.13 | 23.6 | 49.0 | 20.4 | 45.8 | 0.1214 | 5.14 | 3999 | 0.16 | 137 | 2456319 | 2.03 |
| 55 | | ✓ | gamma Cep b | 116727 | 3.21 | 1.03 | 2.13 | 14.102 | 1.26 | 23.7 | 77.6 | 60.1 | 64.7 | 0.1572 | 1.98 | 905.6 | 0.12 | 49.6 | 2453121 | 1.52 |

Notes: Check marks indicate whether the planet is possibly observable. *I* magnitudes were generated from *V* and *B–V* using Neill Reid's photometric data, as described in §2.6 of Brown (2015).

### 3.3. EXPOSURE TIME, τ

For a host star of $I$ magnitude $I_{star}$, we seek the exposure time $\tau$ necessary to achieve the desired value of S/N on a hypothetical planet of magnitude $I_{planet} = I_{star} + \Delta mag$, where $\Delta mag$ is the flux ratio of planet to star expressed in stellar magnitudes:

$$\Delta mag = -2.5 \log \left( p \, \Phi(\beta) \frac{R^2}{x_0^2 + y_0^2} \right), \quad (20)$$

where $p$ is the geometric albedo of the planet, $\Phi$ is the phase function, $\beta$ is the phase angle, which is the planetocentric angle between observer and star:

$$\beta \equiv \arccos \frac{-\sin i \left( x_0 \sin \omega_p + y_0 \cos \omega_p \right)}{\sqrt{x_0^2 + y_0^2}}. \quad (21)$$

and $R$ is the assumed planetary radius in arcseconds.

§2.10 in Brown (2015) states and explains our basic algorithm for computing $\tau$, with two exceptions. First, we use $D'$ (exit aperture) instead of $D$ (entrance aperture) in Equation (15) of Brown (2015), for reasons explained in the third paragraph of §3.1 of this paper. Second, we upgrade the algorithm to allow for a *tapered* central obscuration, as follows.

The inner working angle (*IWA*) is still the nominal radius of the central field obscuration, but the end-to-end efficiency ($h$) is not, however, expected to be a step function at apparent separation $s = IWA$. Due to the finite size of the image—full width at half maximum $FWHM = \lambda / D'$—$h$ is better described as a function of $s$. For sufficiently large separations, $s > s_1$, then $h = h_0$, as in earlier treatments, where

$$h_0 \equiv \varepsilon \, LMC \, reflTrans, \quad (22)$$

where the *LMC* is the Lyot mask correction introduced in §3.1, $\varepsilon$ is the detector quantum efficiency, and *reflTrans* is the product of the reflective or transmissive efficiencies of all the optical elements in series. For sufficiently small separations, $s < s_0$, $h = 0$. Therefore, our model for the dependence of $h$ on $s$ is:

$$h \equiv h_0 \, \mathcal{H}(s), \quad (23)$$

where $\mathcal{H}$ is a piecewise continuous function.

For internal coronagraphs, we assume $h = 0.5$ at $s = IWA$, where the point-spread function is centered on the edge of the obscuration. The exact value of $h$ at $s = IWA$ depends of the design of coronagraph. At smaller angles, $s < IWA$, internally scattered starlight $\zeta$ increases and efficiency $h$ decreases—both so rapidly that we assume $\mathcal{H}(s <$

*IWA*) = 0 for coronagraphs. So far, tolerancing studies for internal coronagraphs have treated only the range $s > IWA$ (Shaklan 2011). Following the literature, then, our taper function for coronagraphs is:

$$\begin{aligned}\mathcal{H}(s) &= 0 & \text{for } s < IWA, \\ &= 0.5\left(1 + \frac{s - IWA}{FWHM}\right) & \text{for } IWA \leq s < IWA + FWHM, \text{ and} \\ &= 1.0 & \text{for } s \geq IWA + FWHM.\end{aligned} \quad (24)$$

A star shade has a solid inner disk surrounded by tapered petals. Here, we define *IWA* to be the angular subtense of the radius of the disk plus the length of the petals, and we use the value *IWA* = 100 mas for both *EXO-S* and *WFIRST-S*.

While some star-shade studies have considered performance for $s < IWA$ (Shaklan 2010, Shaklan 2011a; Glassman et al 2010), we focus in this paper on the *EXO-S* and *WFIRST-S* implementations under study at NASA, which have been toleranced assuming planets are not detectable if $s < IWA$ (Shaklan 2014 private communication). Therefore, we assume a step function for $\mathcal{H}$ in the case of star shades:

$$\begin{aligned}\mathcal{H}(s) &= 0 & \text{for } s < IWA, \text{ and} \\ &= 1.0 & \text{for } s \geq IWA.\end{aligned} \quad (25)$$

## 4. CUBE OF INTERMEDIATE INFORMATION

The earlier sections of this paper prepare us to populate a "cube of intermediate information" that fully realizes a mission for current purposes. Derivatives from the cube—we call them "reports"—can then help construct DRMs, which estimate mission performance, and provide insights into the true nature of the challenges involved with characterizing RV planets.

Our cube has three dimensions: absolute time ($t$ in Julian day), RV planet identifier (HIP or other catalog name), and inclination angle ($i$). $t$ runs from mission start to mission stop, for a total of 1826 days for the assumed five-year mission. RV planets number 16 or 55, for coronagraphs or star shades, respectively, as explained in §3.2. The number of random values of $i$ is 4000 in the current study, generated by the appropriate random deviate, $\cos^{-1}(1-\mathbb{R})$, where $\mathbb{R}$ is a uniform random deviate on the range 0–1. Multiplying the three cardinalities, we see that the cube has 117 million or 402 million cells or addresses, for coronagraphs or for star shades, respectively.

The information in each cell of the cube is a vector with at least eight elements (but possibly more for convenience): (S/N, $s$, $\Delta mag$, $\tau$), plus four flags: $f_{\text{pointing}}$, $f_{\text{SVP\&O}}$, $f_{\text{spatial}}$, and $f_{\text{temporal}}$. For each cell address ($t$, HIP, $i$), and using the parameters in Table 2 that are appropriate to a particular mission, we compute the orbital solution as described in §2.1, compute S/N as described in Equations (18–19), take $s$ from the orbital solution, compute $\Delta mag$ from Equations (20–21), and compute $\tau$ as described in §3.3.

The four flags are binary variables. A "one" means that the flag criterion is satisfied—for that planet, on that day, for that value of $i$—and a "zero" means that the flag criterion is violated. $f_{\text{pointing}}$ refers to solar and anti-solar avoidance ($\gamma_1, \gamma_2$), which is described in detail in §2.8 of Brown (2015). $f_{\text{SVP\&O}}$ is the "single-visit photometric and obscurational" (SVP&O) flag (Brown 2004, Brown 2005). $f_{\text{SVP\&O}} = 1$ if $s >$ IWA and $\Delta mag < \Delta mag_0$, where $\Delta mag_0$ is the limiting delta magnitude, which is the systematic limit or "noise floor," as determined by the stability of the optical system (Brown 2005; Brown 2015). $f_{\text{spatial}}$ pertains to image smearing; it is zero for a planet that would moves more than $FWHM = \lambda / D'$ during the exposure time, but otherwise one. Also, $f_{\text{spatial}}$ is arbitrarily set equal to one, for cosmetic purposes, if $\tau$ is invalid, for example when $s < IWA$ (and $f_{\text{SVP\&O}} = 0$). $f_{\text{temporal}}$ pertains to the continuity of an exposure; it is zero if $f_{\text{pointing}} = 0$ on any day during a multi-day search exposure, or if the exposure time would extend beyond the end of the mission, but it is one otherwise.

We say that the cube "fully realizes a mission" because it contains every important piece of information with a granularity, here, of one mission day (md). We use the cube to produce two types of report: completeness reports and median reports. The premier application of both types of report is to construct design reference missions, as we demonstrate in §5–6.

# 5. DESIGN REFERENCE MISSIONS

A design reference mission (DRM) is a simulation of a mission inside a computer. The general goals of a DRM are (1) to compare alternative mission concepts on an equal, objective basis, (2) to optimize mission parameters, and (3) to set realistic expectations for a mission's productivity and scientific outcomes, should it be implemented.

A DRM is defined by a variety of instrumental, astronomical, and science-operational parameters, listed with baseline values in Table 2. In the current implementation, the cube of intermediate information described in §4 is the link between Table 2 and the DRM. That is, as described in §5.1–5.3, we draw completeness reports and median reports from the cube in order to construct DRMs. The mission parameters and algorithms are fully embodied by the cube.

It is useful to keep in mind that random variables are involved in simulating exoplanet observations. This means that the DRMs themselves are random variables, and that estimating results usually calls for Monte Carlo experiments.

In §6, we present baseline DRMs for the four missions under study.

## 5.1. SCHEDULING ALGORITHM & MERIT FUNCTION, $\mathcal{Z}$

At the heart of a DRM is the scheduling algorithm, with the duty of selecting the next observation after the current exposure is terminated, on any day in the mission. As a rule, the scheduling algorithm selects the observation with the highest current value of the merit function ($\mathcal{Z}$). Obviously, $\mathcal{Z}$ must have a lot of "moving parts" to account for all the factors involved. Nevertheless, the complexity of $\mathcal{Z}$ is more structured and convenient because of the cube. The goal of this section is to deconstruct $\mathcal{Z}$ into completeness reports and median reports from the cube.

$\mathcal{Z}$ (or "discovery rate" or "benefit-to-cost ratio") is the probability of a detecting an RV planet divided by total time cost of the observation, which is the *estimated* exposure time $\tilde{\tau}$ plus the setup time $t_{\text{setup}}$:

$$\mathcal{Z} \equiv \frac{UC}{\tilde{\tau} + t_{\text{setup}}}, \tag{26}$$

In the case of a single search observation—as here for each RV planet—the probability of observational success is the union completeness ($UC$; Brown 2005, Brown & Soummer 2011, Brown 2015). Success—discovering the RV planet in the direct-search observation, yes or no—is a Bernoulli random variable with probability $UC$. In search programs with revisits, the probability is $UC$ with a Bayesian correction (Brown & Soummer 2010; Brown 2015).

§5.2 describes how we estimate the numerator in Equation (26), $UC$, from the cube by means of a completeness report. §5.3 describes how we estimate the search exposure time

$\tilde{\tau}$, in the denominator of Equation (26) from the cube as a median report. §5.3 also describes how we estimate the exposure time $\tilde{\tau}_{spec}$, for possible follow-on spectroscopy, as a function of medians.

## 5.2 UNION COMPLETENESS, *UC*

As explained in §4, we carry four flags in each cell or cube address—$f_{pointing}, f_{SVP\&O}, f_{spatial}$, and $f_{temporal}$. Each flag indicates whether or not a detection criterion is satisfied at that location in the cube—i.e., at that cube address (*t*, HIP, *i*), which corresponds to a potential observation. A potential observation is "invalid"—infeasible or doomed to failure—if any of the four flags is zero, and the observation is "valid" if the flags are all equal to one. For any selected planet, HIP, on any selected mission day, *t*, *UC* is estimated by the fraction of random values of *i* that result in valid cube addresses.

The same procedure works *separately* for each of the four components of UC. That is, we can estimate each constituent completeness—pointing permission, single-visit photometric and obscurational, spatial, or temporal—by the sum of the associated flag divided by the number of trials, which is 4000 here. (*UC* may not, however, be the product of the four constituent *completenesses* because the flags may not be independent.)

See Figures 4 and 6 for two illustrative completeness panels, for mu Ara b observed by EXO-S and for 14 Her b observed by WFIRST-C.

## 5.3  MEDIAN EXPOSURE TIMES, $\tilde{\tau}$ and $\tilde{\tau}_{spec}$

At the time the scheduling algorithm is selecting the next search observation, the actual value of *i* is not known, and therefore the correct exposure time is unknowable in advance. Nevertheless, for each planet and mission day, the cube carries 4000 random values of *i*, each with an associated value of τ. Therefore, we estimate τ as the median of the values of τ *in valid observations* ($\tilde{\tau}$).

$\tilde{\tau}$ is our first example of a median report. By the same principles, we can produce median reports for the other cell contents—S/N, *s*, and Δ*mag,* as well as *functions* of medians, notably in the cases of merit function $\mathcal{Z}$ and the exposure time for follow-on spectroscopy $\tau_{spec}$. These we can estimate by the function of the medians, if we can establish that the function commutes with the median function, which we can in the case of the exposure time calculator and therefore in the denominator of Equation (26).

See Figures 5 and 7 for two illustrative median panels, for mu Ara b observed by EXO-S and for 14 Her b observed by WFIRST-C.

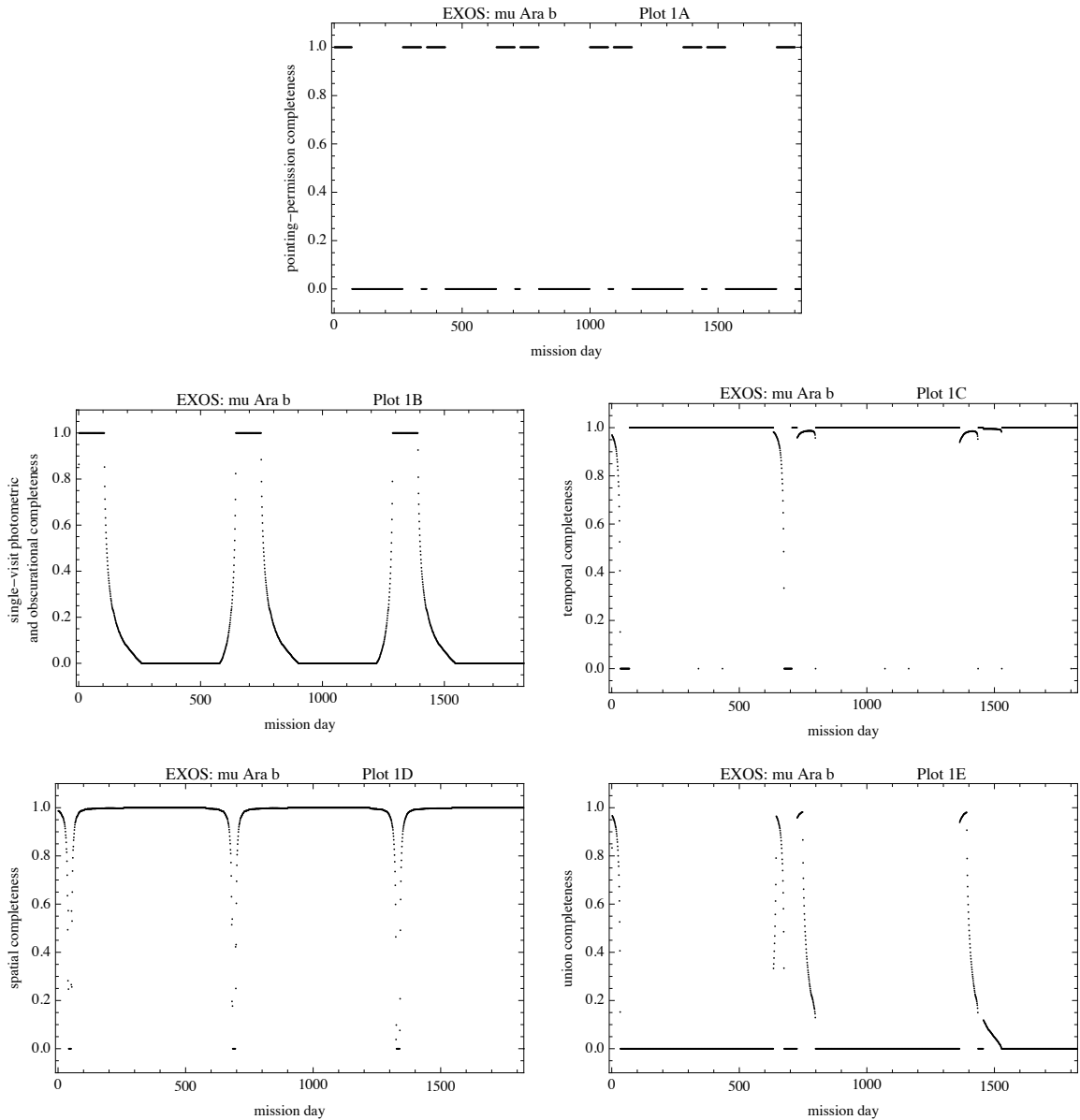

**Figure 4.** Completeness panel for mu Ara b observed by *EXO-S*. Compare with the median panel in Figure 5. Plot 1A shows the doubled windowing due to *both* solar and antisolar avoidance for *EXO-S*. Plot 1B shows orbital period of mu Ara b, 643 days. Plots 1C and 1D show only the first set of node crossings; the second set is absent because those node crossings occurs when $s < IWA$, which means that $\tau$ is invalid due to the SVP&O flag = 0. (See the discussion in the penultimate paragraph of §4.) In Plot 1D, we see that temporal completeness is modulated by a factor not present in spatial completeness: pointing permission, which varies periodically in absolute time (i.e., the abscissa), with a period of the *Earth's* orbital period of 365 days. In Plot 1E, we see that the four-flag *UC* is sparse and spikey, and that opportunities to launch a search observation with high probability of success are only four at most.

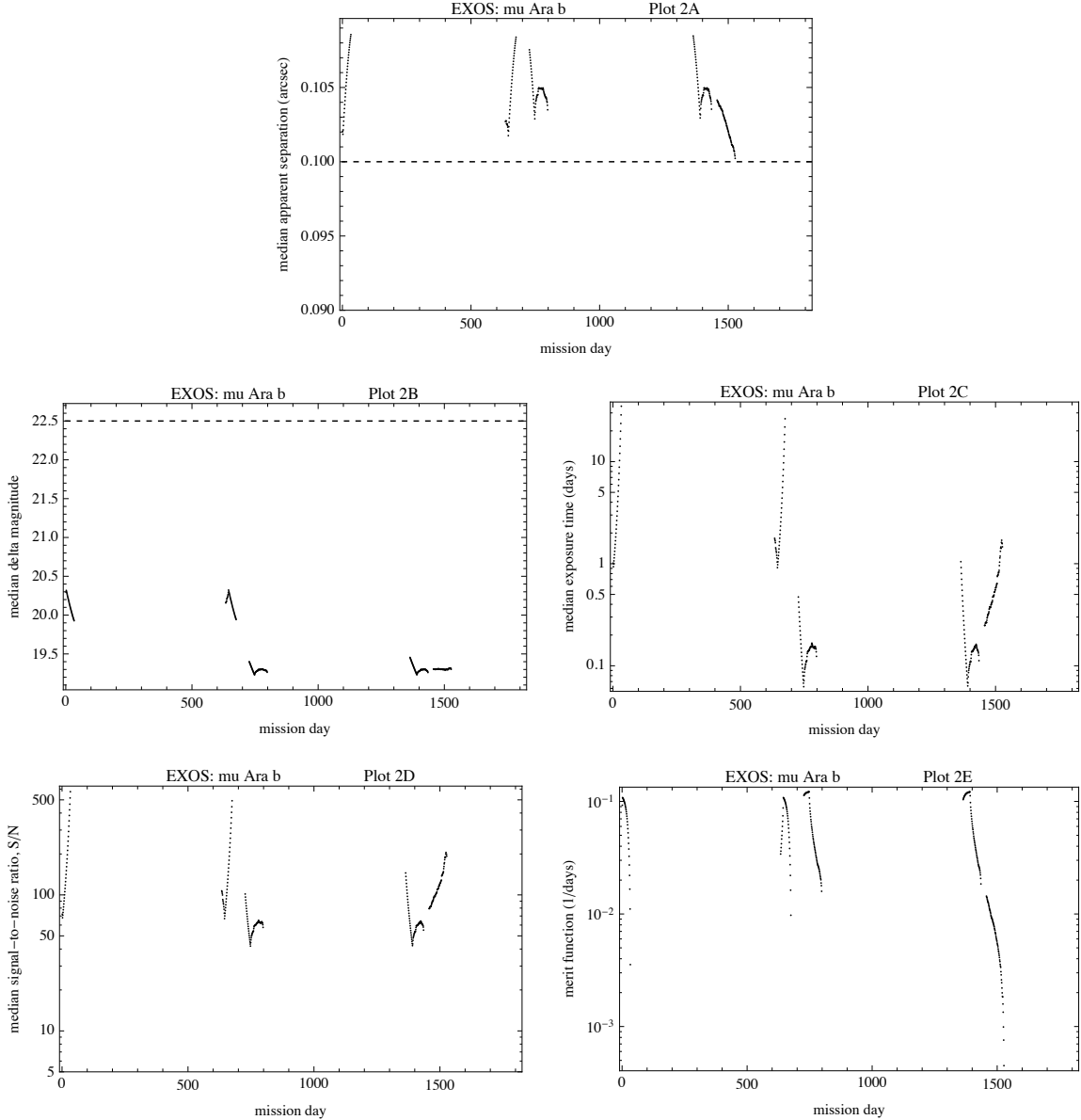

**Figure 5.** Typical median values versus mission day, here for mu Ara b observed by *EXO-S*. Compare with the completeness panel Figure 4. Dashed lines in Plots 2A and 2B: *IWA* and $\Delta mag_0$, which reveal that the regions of value zero in Plot 1B are due to obscuration ($s < IWA$), not to insufficient planet brightness, $\Delta mag > \Delta mag_0$. Plot 2D shows that all valid days (i.e., the four flags are all ones) have S/N requirements elevated above the default S/N = 8, due to their proximity to nodes. In Plot 2E, for merit function $\mathcal{Z}$, notice that the best times to observe mu Ara b (maximum $\mathcal{Z}$) are limited to a few days around 0, 650, and 1350 md, where $\mathcal{Z} \sim 0.1$. In §6, Table 5, line 5, we see that, in developing the design reference mission, the scheduling algorithm chose to observe mu Ara b on mission day 645.

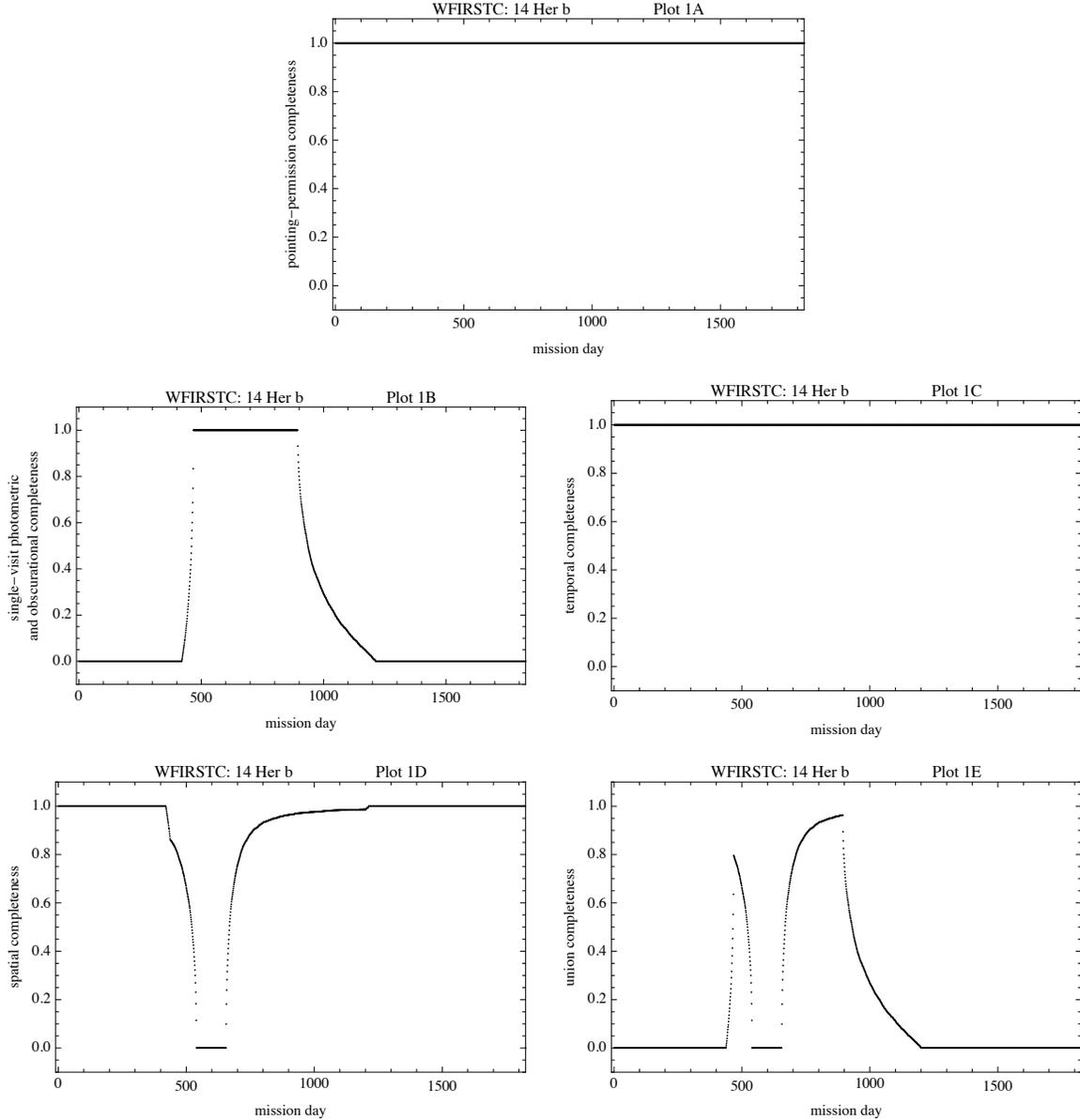

**Figure 6.** Completeness panel for 14 Her b observed by *WFIRST-C*. Plot 1A shows that 14 Her is located in the continuous viewing zone, at ecliptic latitude +63°. Plot 1B shows that the long period of 14 Her b (1773 days) affords only one cycle of SVP&O > 0, in the range 400 < $t$ < 1200 md. Comparing Plot 1B with Plot 1D, we note the divot centered at $t$ = 598 md, which is caused by the sole node crossing by 14 Her b during the five-year mission. From Plot 1E, we see that a search observation with >90% probability of success must be scheduled between 750 and 900 md.

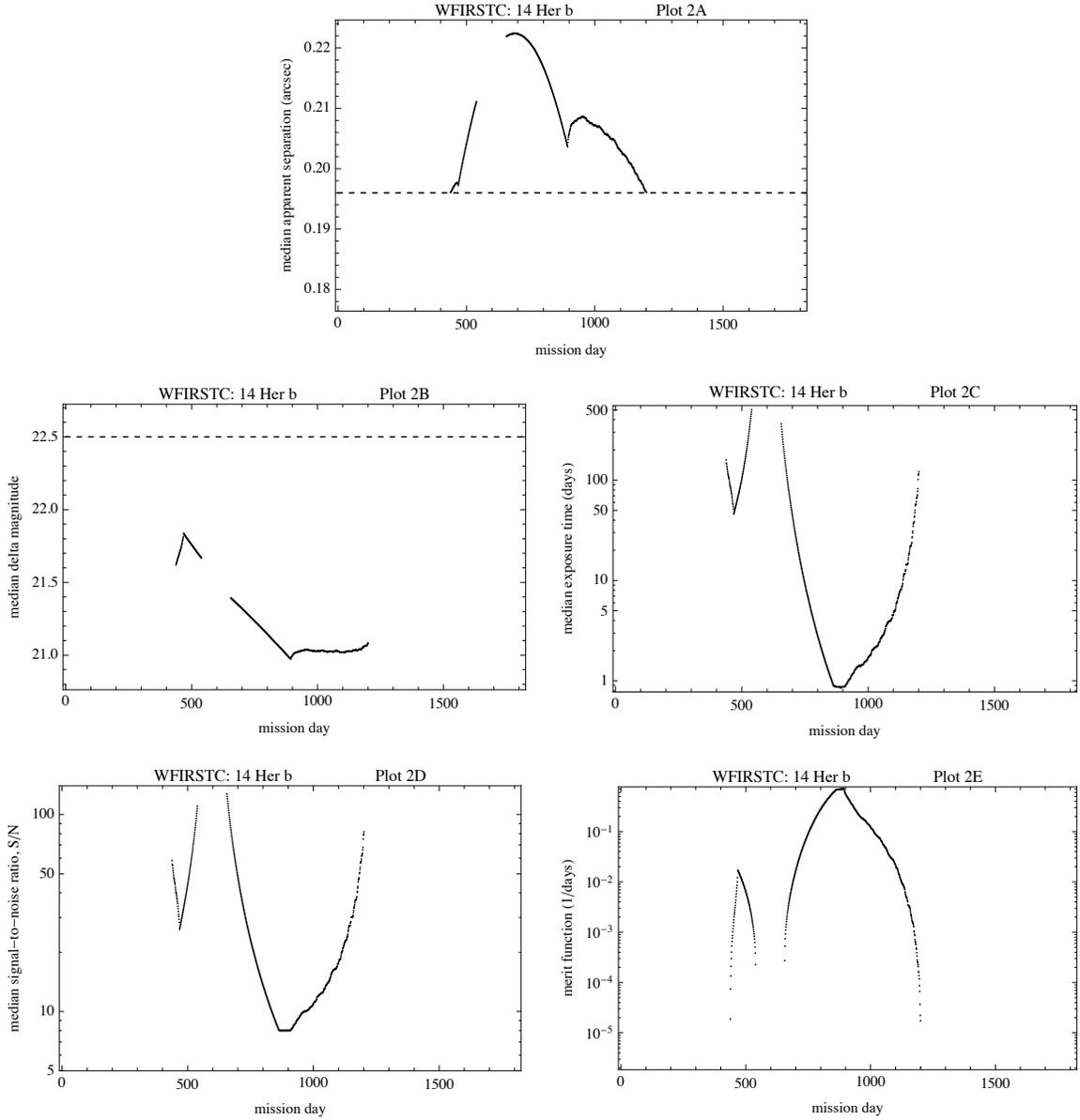

**Figure 7.** Median panel for 14 Her b observed by *WFIRST-C*. Plots 2A and 2B reveal that the regions of value zero for *SVP&OC* in Figure 6, Plot 1B, are due to obscuration ($s <$ *IWA*) not to $\Delta mag > \Delta mag_0$. Plot 2D reveals an elevated S/N requirement (S/N > 8), due to the node crossing at $t = 598$ mission day, everywhere but in a narrow range around $t = 900$ mission days, and that a severe penalty would result if the search mission for 14 Her b were scheduled at different time frame. In §6, Table 6, line 3, we see that the scheduling algorithm chose mission day 894 to observe 14 Her b with *WFIRST-C*, when $\mathcal{Z} \sim 0.71$—consistent with Plot 2E in this figure.

# 6. BASELINE DRMS FOR THE FOUR MISSIONS

The following tables and plots, Tables 4–7 and Figures 8–11, document our four baseline DRMs, for the four missions under study. Each uses the baseline parameters listed for the mission in Table 2. For each observation of a planet, the scheduling algorithm has chosen the day of the mission maximum value of the merit function $\mathcal{Z}$.

For the star-shade missions, *EXO-S* and *WFIRST-S*, $t_{setup}$ is 16-times larger than the 0.5 days for setting up coronagraph missions. Therefore, in most cases, the median exposure time $\tilde{\tau}$ in column 12 is much smaller than the setup time, $t_{setup}$. In this case, Equation (26) says that prioritizing observations by $UC$ is the same as prioritizing by $\tilde{\tau}$. As a result, the star-shade missions may observe fewer RV planets total, but they can be the ones with $UC > 0.90$, say, which means they have guaranteed success at a 90% confidence level.

No attempt has been made to identify scheduling conflicts, which are only a minor issue at worst, for at least three reasons. First, Figures 8–11 indicate that the schedules are dilute. Second, by the time any of these missions are implemented, whatever uncertainty remains in the radial-velocity solutions—which will be far less after ten more years of radial-velocity observations—will afford further scheduling flexibility due to orbit uncertainty, which is not accounted here. Third, the schedule is further diluted when the most unpromising targets (low $\mathcal{Z}$, say $\mathcal{Z} < 0.01$) are removed from consideration, as they are in Figures 8–11. These targets are poor choices, either because they are too costly in observing time—say more than 100 days for a high probability target ($UC \sim 1$), or because the observing time is reasonable (~ 1 day) but the probability of success finding the planet is too low ($UC \sim 0.01$)—or some poor combination in between.

By this argument, we exclude planets with $\mathcal{Z} \leq 0.01$ from the graphical DRMs, but they remain included in the tabular DRMs.

With these caveats in mind, we can proceed in §7 to interpret the results in Tables 4–7 and Figures 8–11, in which observations are listed or displayed in rank order according to $\mathcal{Z}$—mostly the same order as if we ranked by $UC$).

Table 4. Tabular DRM for *WFIRST-S*.

| | planet name | median apparent separation s | median flux ratio delMag | median S/N | mission day | union completeness UC | cumulative UC | observing time (days) | cumulative observing time (days) | merit function (1/days) | median search exposure time | median spectroscopic exposure time (days) |
|---|---|---|---|---|---|---|---|---|---|---|---|---|
| 1 | beta Gem b | 0.137 | 21.2 | 8.0 | 476 | 1.000 | 1.00 | 8.000 | 8.0 | 0.125 | 0.00 | 0.00 |
| 2 | epsilon Eri b | 0.577 | 19.9 | 8.0 | 818 | 1.000 | 2.00 | 8.000 | 16.0 | 0.125 | 0.00 | 0.00 |
| 3 | gamma Cep b | 0.118 | 19.4 | 8.0 | 512 | 0.999 | 3.00 | 8.000 | 24.0 | 0.125 | 0.00 | 0.00 |
| 4 | HD 192310 c | 0.132 | 18.6 | 8.0 | 60 | 0.999 | 4.00 | 8.000 | 32.0 | 0.125 | 0.00 | 0.00 |
| 5 | 47 UMa c | 0.152 | 20.2 | 8.0 | 1719 | 0.999 | 5.00 | 8.001 | 40.0 | 0.125 | 0.00 | 0.02 |
| 6 | upsilon And d | 0.139 | 19.6 | 10.8 | 149 | 0.998 | 5.99 | 8.000 | 48.0 | 0.125 | 0.00 | 0.01 |
| 7 | 47 UMa b | 0.114 | 19.3 | 8.0 | 1253 | 0.998 | 6.99 | 8.000 | 56.0 | 0.125 | 0.00 | 0.01 |
| 8 | HD 190360 b | 0.218 | 21.1 | 8.0 | 1066 | 0.997 | 7.99 | 8.011 | 64.0 | 0.124 | 0.01 | 0.16 |
| 9 | GJ 832 b | 0.342 | 20.0 | 8.0 | 797 | 0.998 | 8.99 | 8.017 | 72.0 | 0.124 | 0.02 | 0.24 |
| 10 | HD 10647 b | 0.107 | 19.6 | 9.3 | 391 | 0.996 | 9.98 | 8.002 | 80.0 | 0.124 | 0.00 | 0.02 |
| 11 | mu Ara b | 0.103 | 19.2 | 20.0 | 748 | 0.996 | 10.98 | 8.003 | 88.0 | 0.124 | 0.00 | 0.04 |
| 12 | mu Ara c | 0.282 | 21.5 | 8.0 | 16 | 0.996 | 11.98 | 8.009 | 96.0 | 0.124 | 0.01 | 0.12 |
| 13 | 7 CMa b | 0.101 | 19.8 | 57.3 | 180 | 0.996 | 12.97 | 8.008 | 104.1 | 0.124 | 0.01 | 0.11 |
| 14 | HD 154345 b | 0.147 | 20.8 | 8.0 | 1350 | 0.996 | 13.97 | 8.039 | 112.1 | 0.124 | 0.04 | 0.55 |
| 15 | GJ 649 b | 0.108 | 18.4 | 8.1 | 1365 | 0.992 | 14.96 | 8.008 | 120.1 | 0.124 | 0.01 | 0.11 |
| 16 | 14 Her b | 0.174 | 20.7 | 8.0 | 1032 | 0.994 | 15.95 | 8.021 | 128.1 | 0.124 | 0.02 | 0.29 |
| 17 | HD 147513 b | 0.104 | 20.4 | 13.6 | 1753 | 0.989 | 16.94 | 8.008 | 136.1 | 0.123 | 0.01 | 0.12 |
| 18 | HD 128311 c | 0.110 | 19.5 | 8.0 | 1796 | 0.987 | 17.93 | 8.011 | 144.1 | 0.123 | 0.01 | 0.15 |
| 19 | HD 217107 c | 0.206 | 21.3 | 8.0 | 1 | 0.989 | 18.92 | 8.036 | 152.2 | 0.123 | 0.04 | 0.50 |
| 20 | HD 134987 c | 0.161 | 21.4 | 8.0 | 1088 | 0.983 | 19.90 | 8.072 | 160.2 | 0.122 | 0.07 | 1.01 |
| 21 | GJ 849 b | 0.173 | 19.4 | 8.0 | 334 | 0.987 | 20.89 | 8.141 | 168.4 | 0.121 | 0.14 | 1.97 |
| 22 | GJ 676 A b | 0.107 | 19.4 | 8.0 | 1733 | 0.969 | 21.86 | 8.074 | 176.5 | 0.120 | 0.07 | 1.03 |
| 23 | HD 87883 b | 0.120 | 19.9 | 12.7 | 1633 | 0.962 | 22.82 | 8.059 | 184.5 | 0.119 | 0.06 | 0.82 |
| 24 | HD 216437 b | 0.101 | 20.4 | 38.7 | 1490 | 0.973 | 23.79 | 8.180 | 192.7 | 0.119 | 0.18 | 2.52 |
| 25 | HD 38529 c | 0.115 | 21.5 | 17.8 | 1688 | 0.954 | 24.74 | 8.155 | 200.9 | 0.117 | 0.16 | 2.17 |
| 26 | HD 39091 b | 0.197 | 21.1 | 36.5 | 40 | 0.926 | 25.67 | 8.282 | 209.1 | 0.112 | 0.28 | 3.95 |
| 27 | HD 70642 b | 0.102 | 20.5 | 21.7 | 1375 | 0.926 | 26.59 | 8.390 | 217.5 | 0.110 | 0.39 | 5.46 |
| 28 | HD 50554 b | 0.102 | 20.7 | 30.1 | 843 | 0.916 | 27.51 | 8.669 | 226.2 | 0.106 | 0.67 | 9.36 |
| 29 | HD 117207 b | 0.108 | 21.0 | 8.9 | 1675 | 0.838 | 28.35 | 8.167 | 234.4 | 0.103 | 0.17 | 2.34 |
| 30 | HD 142022 b | 0.115 | 21.2 | 16.0 | 1003 | 0.939 | 29.29 | 9.542 | 243.9 | 0.098 | 1.54 | 21.59 |
| 31 | GJ 179 b | 0.140 | 19.8 | 8.0 | 214 | 0.944 | 30.23 | 9.919 | 253.8 | 0.095 | 1.92 | 26.86 |
| 32 | HD 169830 c | 0.105 | 22.2 | 30.2 | 1158 | 0.858 | 31.09 | 10.361 | 264.2 | 0.083 | 2.36 | 33.05 |
| 33 | HD 33636 b | 0.139 | 21.3 | 31.7 | 958 | 0.744 | 31.83 | 10.726 | 274.9 | 0.069 | 2.73 | 38.17 |
| 34 | HD 13931 b | 0.107 | 21.6 | 16.5 | 1653 | 0.729 | 32.56 | 11.598 | 286.5 | 0.063 | 3.60 | 50.37 |
| 35 | HD 181433 d | 0.119 | 20.8 | 23.7 | 69 | 0.675 | 33.24 | 11.434 | 297.9 | 0.059 | 3.43 | 48.07 |
| 36 | GJ 317 b | 0.107 | 20.0 | 9.7 | 1328 | 0.908 | 34.14 | 17.619 | 315.6 | 0.052 | 9.52 | 133.24 |
| 37 | HD 30562 b | 0.128 | 20.9 | 8.0 | 798 | 0.385 | 34.53 | 8.011 | 323.6 | 0.048 | 0.01 | 0.15 |
| 38 | 16 Cyg B b | 0.116 | 20.2 | 9.7 | 1799 | 0.251 | 34.78 | 8.010 | 331.6 | 0.031 | 0.01 | 0.14 |
| 39 | HD 89307 b | 0.116 | 22.0 | 44.7 | 1611 | 0.623 | 35.40 | 29.573 | 361.2 | 0.021 | 21.57 | 302.02 |
| 40 | HD 204941 b | 0.114 | 22.0 | 16.5 | 1543 | 0.592 | 35.99 | 32.766 | 393.9 | 0.018 | 24.77 | 346.72 |
| 41 | HD 222155 b | 0.108 | 21.8 | 13.4 | 1618 | 0.159 | 36.15 | 9.527 | 403.4 | 0.017 | 1.53 | 21.38 |
| 42 | HD 196885 b | 0.105 | 21.0 | 26.4 | 828 | 0.109 | 36.26 | 8.460 | 411.9 | 0.013 | 0.46 | 6.44 |
| 43 | HD 7449 b | 0.104 | 22.4 | 24.4 | 742 | 0.432 | 36.70 | 43.382 | 455.3 | 0.010 | 35.38 | 495.35 |
| 44 | 55 Cnc d | 0.442 | 22.5 | 72.0 | 1725 | 0.028 | 36.72 | 20.054 | 475.3 | 0.001 | 12.05 | 168.76 |
| 45 | kappa CrB b | 0.100 | 20.9 | 867.3 | 731 | 0.000 | 36.72 | 26.825 | 502.2 | 0.000 | 18.82 | 263.55 |
| 46 | HD 24040 b | ... | ... | ... | ... | ... | ... | ... | ... | ... | ... | ... |
| 47 | HD 72659 b | ... | ... | ... | ... | ... | ... | ... | ... | ... | ... | ... |
| 48 | GJ 328 b | ... | ... | ... | ... | ... | ... | ... | ... | ... | ... | ... |
| 49 | HD 204313 d | ... | ... | ... | ... | ... | ... | ... | ... | ... | ... | ... |
| 50 | HD 79498 b | ... | ... | ... | ... | ... | ... | ... | ... | ... | ... | ... |
| 51 | HD 220773 b | ... | ... | ... | ... | ... | ... | ... | ... | ... | ... | ... |
| 52 | HD 50499 b | ... | ... | ... | ... | ... | ... | ... | ... | ... | ... | ... |
| 53 | HD 10180 h | ... | ... | ... | ... | ... | ... | ... | ... | ... | ... | ... |
| 54 | HD 187123 c | ... | ... | ... | ... | ... | ... | ... | ... | ... | ... | ... |
| 55 | HD 106252 b | ... | ... | ... | ... | ... | ... | ... | ... | ... | ... | ... |

Notes: *UC* is the probability of a successful detection. The scheduling algorithm chooses the mission day of maximize $\mathcal{Z}$. Cumulative *UC* is the expectation value of the total number of planets recovered, to that point. The numbers labeling the points in Figure 6 refer to line numbers in this table. The red points in Figure 6 are "guaranteed" because they offer a 90% or better probability of successfully achieving the mass accuracy in Equation (1). Ellipses indicate that a planet is not observable at all, assuming the parameters in Table 2.

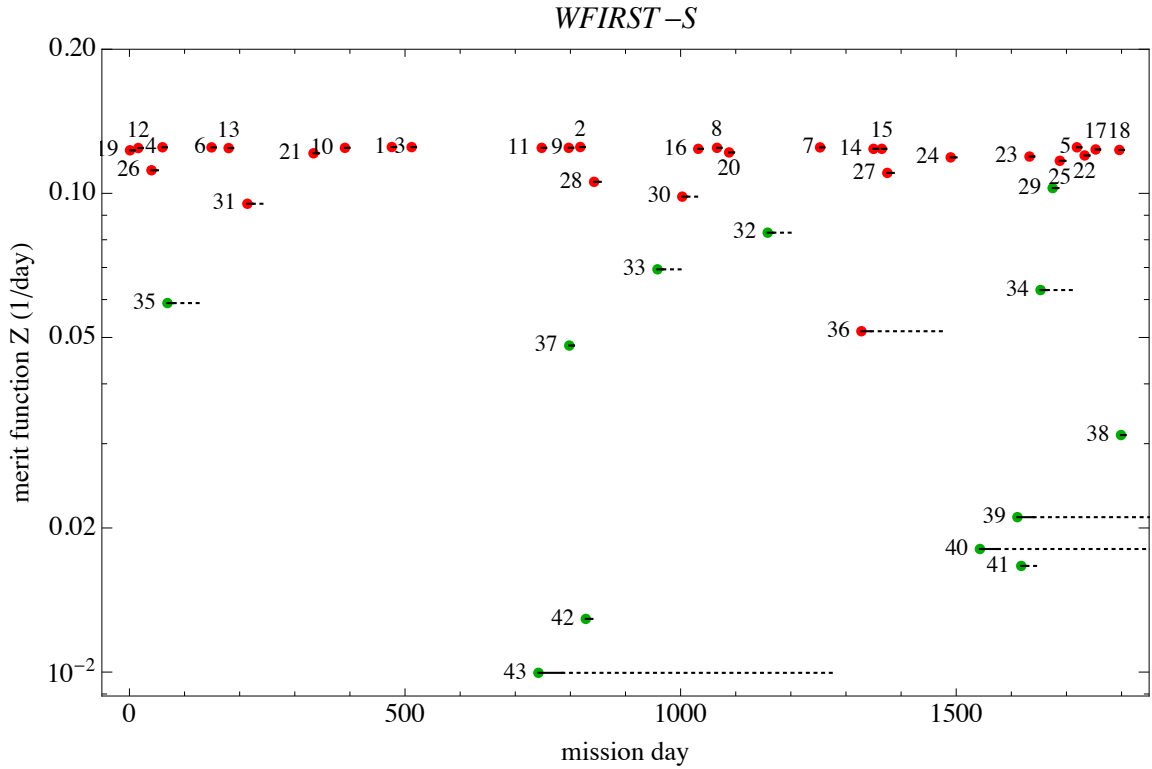

**Figure 8.** Graphical DRM for *WFIRST-S*. Numbered dots mark the value of the median merit function $\mathcal{Z}$ and the mission day of the search exposure with maximum $\mathcal{Z}$, as selected by the scheduling algorithm. Red: targets with "guaranteed" success in mass determination, due to $UC \geq 0.9$. Green: riskier targets with $UC < 0.9$. The labeling numbers correspond to the row numbers in Table 4. A solid line, when it extends beyond the dot, gives the median exposure plus setup time of the search observation. Dashed lines show the median exposure time for the possible follow-on spectroscopic exposure. Long spectroscopic exposures may not be tolerated, for example, by the pointing restrictions. In this plot, candidate observations are cut off at $\mathcal{Z}_{\min} = 0.01$, but all results are reported in Table 4.

Table 5. Tabular DRM for *EXO-S*.

| | planet name | median apparent separation s | median flux ratio delMag | median S/N | mission day | union completeness UC | cumulative UC | observing time (days) | cumulative observing time (days) | merit function (1/days) | median search exposure time | median spectroscopic exposure time (days) |
|---|---|---|---|---|---|---|---|---|---|---|---|---|
| 1 | beta Gem b | 0.137 | 21.2 | 10.4 | 476 | 1.000 | 1.00 | 8.000 | 8.0 | 0.125 | 0.00 | 0.01 |
| 2 | epsilon Eri b | 0.560 | 19.9 | 8.0 | 812 | 1.000 | 2.00 | 8.001 | 16.0 | 0.125 | 0.00 | 0.01 |
| 3 | gamma Cep b | 0.117 | 19.4 | 8.0 | 512 | 0.999 | 3.00 | 8.000 | 24.0 | 0.125 | 0.00 | 0.00 |
| 4 | upsilon And d | 0.126 | 19.4 | 13.9 | 128 | 0.995 | 3.99 | 8.003 | 32.0 | 0.124 | 0.00 | 0.05 |
| 5 | 47 UMa b | 0.113 | 19.3 | 9.8 | 1253 | 0.995 | 4.99 | 8.004 | 40.0 | 0.124 | 0.00 | 0.08 |
| 6 | HD 192310 c | 0.127 | 18.6 | 8.0 | 49 | 0.995 | 5.98 | 8.002 | 48.0 | 0.124 | 0.00 | 0.04 |
| 7 | 47 UMa c | 0.143 | 20.1 | 8.0 | 1701 | 0.994 | 6.98 | 8.009 | 56.0 | 0.124 | 0.01 | 0.22 |
| 8 | HD 10647 b | 0.106 | 19.6 | 19.7 | 391 | 0.987 | 7.96 | 8.053 | 64.1 | 0.123 | 0.05 | 1.27 |
| 9 | HD 190360 b | 0.197 | 21.0 | 8.0 | 1124 | 0.987 | 8.95 | 8.100 | 72.2 | 0.122 | 0.10 | 2.60 |
| 10 | mu Ara b | 0.103 | 19.2 | 42.1 | 748 | 0.982 | 9.93 | 8.062 | 80.2 | 0.122 | 0.06 | 1.37 |
| 11 | mu Ara c | 0.277 | 21.4 | 8.0 | 1 | 0.985 | 10.92 | 8.090 | 88.3 | 0.122 | 0.09 | 2.30 |
| 12 | GJ 832 b | 0.346 | 19.9 | 8.0 | 1041 | 0.993 | 11.91 | 8.179 | 96.5 | 0.121 | 0.18 | 4.74 |
| 13 | 7 CMa b | 0.101 | 19.8 | 120.8 | 180 | 0.982 | 12.89 | 8.142 | 104.6 | 0.121 | 0.14 | 2.67 |
| 14 | 14 Her b | 0.169 | 20.7 | 8.0 | 1043 | 0.962 | 13.85 | 8.223 | 112.9 | 0.117 | 0.22 | 5.89 |
| 15 | HD 147513 b | 0.104 | 20.4 | 28.6 | 1753 | 0.966 | 14.82 | 8.339 | 121.2 | 0.116 | 0.34 | 8.49 |
| 16 | HD 154345 b | 0.143 | 20.8 | 8.0 | 1350 | 0.975 | 15.80 | 8.461 | 129.7 | 0.115 | 0.46 | 12.27 |
| 17 | GJ 649 b | 0.107 | 18.4 | 16.8 | 1365 | 0.962 | 16.76 | 8.351 | 138.0 | 0.115 | 0.35 | 9.19 |
| 18 | HD 217107 c | 0.203 | 21.3 | 8.0 | 1 | 0.936 | 17.69 | 8.416 | 146.4 | 0.111 | 0.42 | 11.01 |
| 19 | HD 128311 c | 0.107 | 19.5 | 14.4 | 1803 | 0.916 | 18.61 | 8.339 | 154.8 | 0.110 | 0.34 | 8.89 |
| 20 | HD 134987 c | 0.158 | 21.4 | 8.0 | 1087 | 0.897 | 19.51 | 8.856 | 163.6 | 0.101 | 0.86 | 22.83 |
| 21 | GJ 849 b | 0.168 | 19.4 | 8.0 | 334 | 0.925 | 20.43 | 9.720 | 173.4 | 0.095 | 1.72 | 46.43 |
| 22 | HD 87883 b | 0.110 | 19.7 | 19.3 | 1650 | 0.801 | 21.23 | 9.139 | 182.5 | 0.088 | 1.14 | 30.17 |
| 23 | GJ 676 A b | 0.105 | 19.4 | 15.6 | 1733 | 0.846 | 22.08 | 11.316 | 193.8 | 0.075 | 3.32 | 89.16 |
| 24 | HD 216437 b | 0.101 | 20.4 | 76.4 | 1490 | 0.827 | 22.91 | 15.057 | 208.9 | 0.055 | 7.06 | 182.75 |
| 25 | HD 38529 c | 0.113 | 21.4 | 31.9 | 1693 | 0.678 | 23.58 | 13.433 | 222.3 | 0.050 | 5.43 | 142.98 |
| 26 | HD 117207 b | 0.106 | 20.9 | 16.4 | 1675 | 0.677 | 24.26 | 14.632 | 236.9 | 0.046 | 6.63 | 178.28 |
| 27 | HD 30562 b | 0.128 | 20.9 | 11.5 | 798 | 0.352 | 24.61 | 8.241 | 245.2 | 0.043 | 0.24 | 6.26 |
| 28 | 16 Cyg B b | 0.115 | 20.1 | 21.9 | 1067 | 0.222 | 24.83 | 8.506 | 253.7 | 0.026 | 0.51 | 13.10 |
| 29 | HD 39091 b | 0.197 | 21.1 | 66.7 | 38 | 0.446 | 25.28 | 18.094 | 271.8 | 0.025 | 10.09 | 262.71 |
| 30 | HD 70642 b | 0.101 | 20.5 | 38.5 | 1375 | 0.503 | 25.78 | 21.850 | 293.6 | 0.023 | 13.85 | 370.28 |
| 31 | GJ 179 b | 0.133 | 19.8 | 8.0 | 200 | 0.599 | 26.38 | 32.146 | 325.8 | 0.019 | 24.15 | 656.06 |
| 32 | HD 50554 b | 0.100 | 20.6 | 53.9 | 843 | 0.481 | 26.86 | 32.134 | 357.9 | 0.015 | 24.13 | 643.96 |
| 33 | HD 33636 b | 0.105 | 20.6 | 35.7 | 1171 | 0.203 | 27.07 | 20.842 | 378.7 | 0.010 | 12.84 | 343.40 |
| 34 | HD 142022 b | 0.113 | 21.2 | 29.1 | 1003 | 0.601 | 27.67 | 69.776 | 448.5 | 0.009 | 61.78 | 1668.75 |
| 35 | HD 196885 b | 0.102 | 20.9 | 42.5 | 812 | 0.058 | 27.72 | 20.575 | 469.1 | 0.003 | 12.58 | 334.24 |
| 36 | GJ 317 b | 0.103 | 18.1 | 39.5 | 141 | 0.117 | 27.84 | 73.143 | 542.2 | 0.002 | 65.14 | 1762.61 |
| 37 | HD 222155 b | 0.103 | 21.7 | 22.3 | 1545 | 0.066 | 27.91 | 55.196 | 597.4 | 0.001 | 47.20 | 1273.53 |
| 38 | HD 24040 b | ... | ... | ... | ... | ... | ... | ... | ... | ... | ... | ... |
| 39 | HD 72659 b | ... | ... | ... | ... | ... | ... | ... | ... | ... | ... | ... |
| 40 | GJ 328 b | ... | ... | ... | ... | ... | ... | ... | ... | ... | ... | ... |
| 41 | HD 204313 d | ... | ... | ... | ... | ... | ... | ... | ... | ... | ... | ... |
| 42 | HD 79498 b | ... | ... | ... | ... | ... | ... | ... | ... | ... | ... | ... |
| 43 | HD 13931 b | ... | ... | ... | ... | ... | ... | ... | ... | ... | ... | ... |
| 44 | HD 7449 b | ... | ... | ... | ... | ... | ... | ... | ... | ... | ... | ... |
| 45 | HD 220773 b | ... | ... | ... | ... | ... | ... | ... | ... | ... | ... | ... |
| 46 | HD 181433 d | ... | ... | ... | ... | ... | ... | ... | ... | ... | ... | ... |
| 47 | kappa CrB b | ... | ... | ... | ... | ... | ... | ... | ... | ... | ... | ... |
| 48 | HD 204941 b | ... | ... | ... | ... | ... | ... | ... | ... | ... | ... | ... |
| 49 | HD 50499 b | ... | ... | ... | ... | ... | ... | ... | ... | ... | ... | ... |
| 50 | HD 89307 b | ... | ... | ... | ... | ... | ... | ... | ... | ... | ... | ... |
| 51 | HD 169830 c | ... | ... | ... | ... | ... | ... | ... | ... | ... | ... | ... |
| 52 | 55 Cnc d | ... | ... | ... | ... | ... | ... | ... | ... | ... | ... | ... |
| 53 | HD 10180 h | ... | ... | ... | ... | ... | ... | ... | ... | ... | ... | ... |
| 54 | HD 187123 c | ... | ... | ... | ... | ... | ... | ... | ... | ... | ... | ... |
| 55 | HD 106252 b | ... | ... | ... | ... | ... | ... | ... | ... | ... | ... | ... |

Notes: See notes for Table 4.

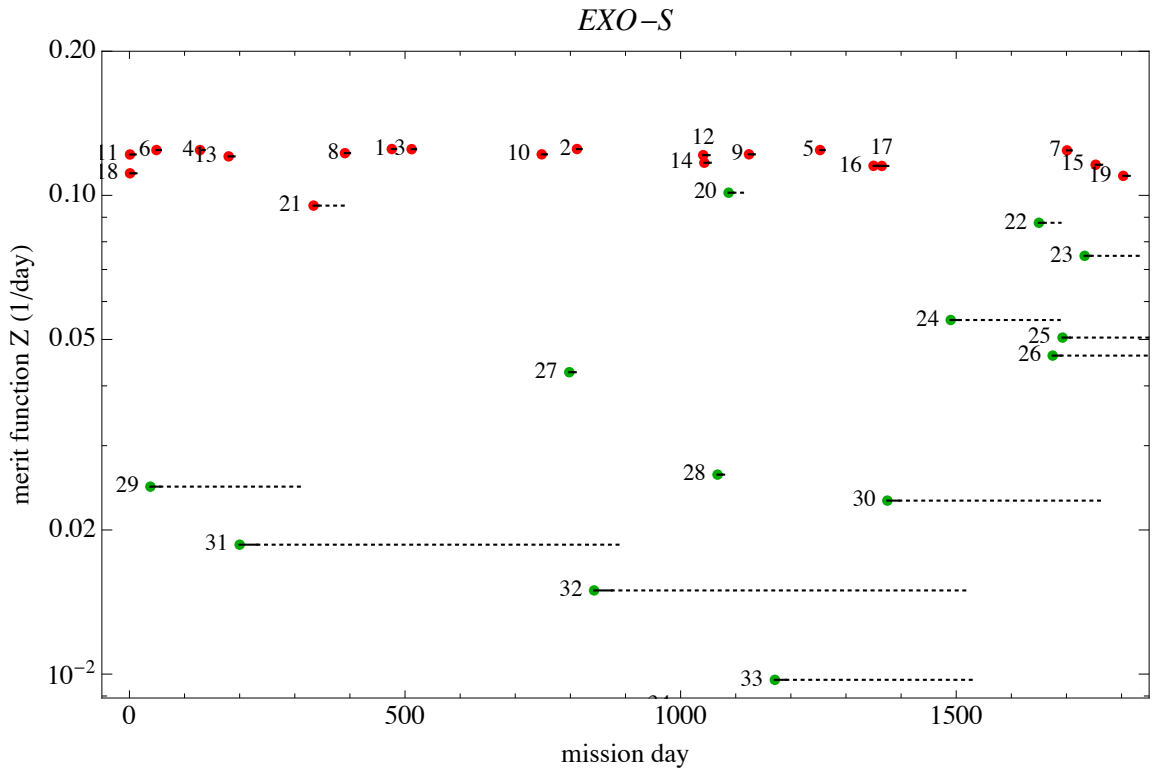

**Figure 9.** Graphical DRM for *EXO-S*. See the caption of Figure 8.

Table 6. Tabular DRM for *WFIRST-C*.

| | planet name | median apparent separation s | median flux ratio delMag | median S/N | mission day | union completeness UC | cumulative UC | observing time (days) | cumulative observing time (days) | merit function (1/days) | median search exposure time | median spectroscopic exposure time (days) |
|---|---|---|---|---|---|---|---|---|---|---|---|---|
| 1 | epsilon Eri b | 0.543 | 19.9 | 8.0 | 797 | 0.999 | 1.00 | 0.503 | 0.5 | 1.988 | 0.00 | 0.04 |
| 2 | 47 UMa c | 0.217 | 20.6 | 8.0 | 1227 | 0.986 | 1.99 | 0.563 | 1.1 | 1.753 | 0.06 | 0.88 |
| 3 | mu Ara c | 0.283 | 21.5 | 8.0 | 35 | 0.989 | 2.97 | 0.661 | 1.7 | 1.496 | 0.16 | 2.26 |
| 4 | GJ 832 b | 0.360 | 19.9 | 8.0 | 1069 | 0.984 | 3.96 | 0.722 | 2.4 | 1.362 | 0.22 | 3.11 |
| 5 | HD 190360 b | 0.241 | 21.2 | 8.0 | 972 | 0.989 | 4.95 | 0.785 | 3.2 | 1.259 | 0.27 | 3.74 |
| 6 | 14 Her b | 0.204 | 21.0 | 8.0 | 894 | 0.963 | 5.91 | 1.363 | 4.6 | 0.707 | 0.86 | 12.08 |
| 7 | HD 217107 c | 0.242 | 21.6 | 8.0 | 120 | 0.959 | 6.87 | 1.582 | 6.2 | 0.606 | 1.08 | 15.15 |
| 8 | HD 154345 b | 0.203 | 21.1 | 8.0 | 1676 | 0.949 | 7.82 | 2.222 | 8.4 | 0.427 | 1.72 | 24.10 |
| 9 | upsilon And d | 0.214 | 21.9 | 8.9 | 533 | 0.177 | 7.99 | 0.839 | 9.2 | 0.210 | 0.21 | 2.99 |
| 10 | GJ 849 b | 0.225 | 19.8 | 8.0 | 1766 | 0.924 | 8.92 | 4.968 | 14.2 | 0.186 | 4.47 | 62.52 |
| 11 | HD 39091 b | 0.197 | 21.1 | 35.4 | 39 | 0.869 | 9.79 | 8.794 | 23.0 | 0.099 | 8.29 | 116.12 |
| 12 | HD 134987 c | 0.200 | 21.8 | 11.4 | 1693 | 0.047 | 9.83 | 9.362 | 32.4 | 0.005 | 8.86 | 124.07 |
| 13 | GJ 328 b | ... | ... | ... | ... | ... | ... | ... | ... | ... | ... | ... |
| 14 | 55 Cnc d | ... | ... | ... | ... | ... | ... | ... | ... | ... | ... | ... |
| 15 | GJ 179 b | ... | ... | ... | ... | ... | ... | ... | ... | ... | ... | ... |
| 16 | HD 87883 b | ... | ... | ... | ... | ... | ... | ... | ... | ... | ... | ... |

Notes: See notes for Table 4.

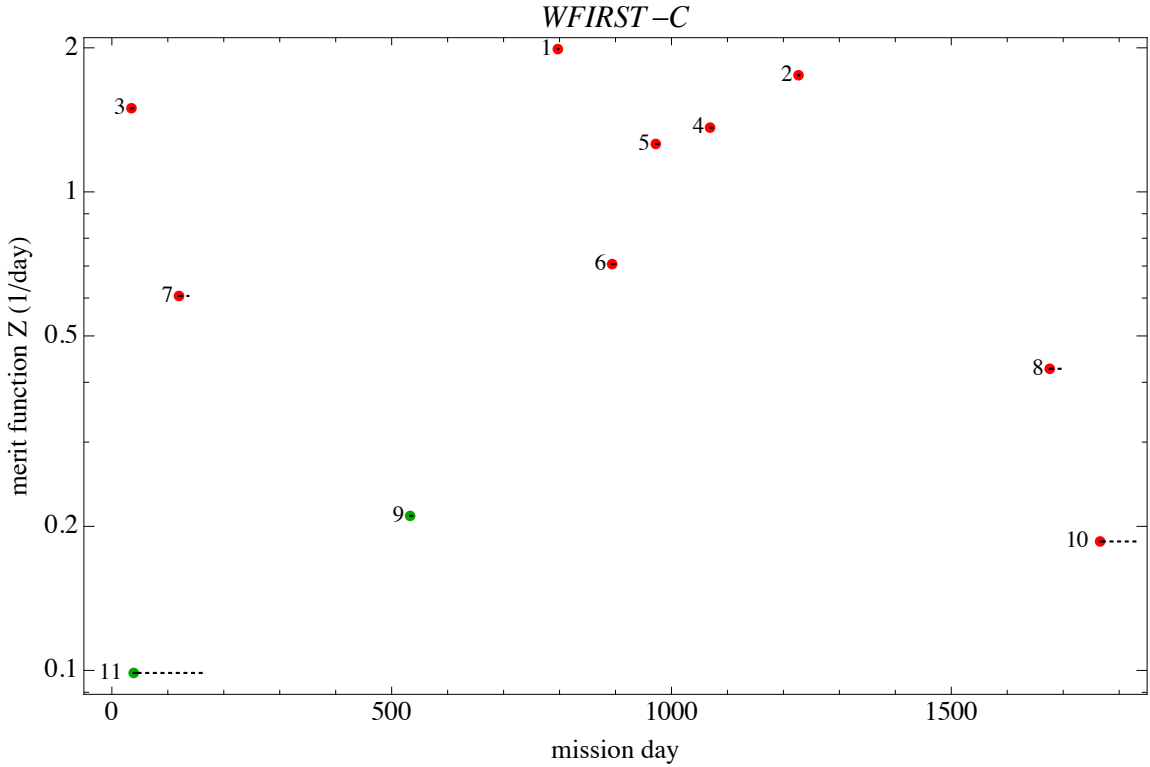

**Figure 10:** Graphical DRM for *WFIRST-C*. See the caption for Figure 8.

Table 7. Tabular DRM for *EXO-C*.

| | planet name | median apparent separation s | median flux ratio delMag | median S/N | mission day | union completeness UC | cumulative UC | observing time (days) | cumulative observing time (days) | merit function (1/days) | median search exposure time | median spectroscopic exposure time (days) |
|---|---|---|---|---|---|---|---|---|---|---|---|---|
| 1 | epsilon Eri b | 0.533 | 19.9 | 8.0 | 790 | 0.999 | 1.00 | 0.502 | 0.5 | 1.991 | 0.00 | 0.04 |
| 2 | mu Ara c | 0.280 | 21.4 | 8.0 | 23 | 0.980 | 1.98 | 0.890 | 1.4 | 1.101 | 0.39 | 12.10 |
| 3 | GJ 832 b | 0.420 | 20.0 | 8.0 | 1157 | 0.981 | 2.96 | 1.166 | 2.6 | 0.841 | 0.67 | 18.76 |
| 4 | HD 190360 b | 0.250 | 21.2 | 8.0 | 938 | 0.975 | 3.94 | 1.420 | 4.0 | 0.687 | 0.90 | 31.72 |
| 5 | 47 UMa c | 0.230 | 20.8 | 21.3 | 1183 | 0.942 | 4.88 | 1.659 | 5.6 | 0.568 | 1.16 | 42.21 |
| 6 | HD 217107 c | 0.238 | 21.6 | 10.0 | 111 | 0.914 | 5.79 | 6.940 | 12.6 | 0.132 | 6.44 | 242.07 |
| 7 | GJ 849 b | 0.228 | 19.8 | 10.9 | 1742 | 0.563 | 6.35 | 41.696 | 54.3 | 0.014 | 41.20 | 1660.23 |
| 8 | upsilon And d | 0.230 | 21.6 | 63.4 | 660 | 0.051 | 6.40 | 11.489 | 65.8 | 0.004 | 10.99 | 381.79 |
| 9 | HD 154345 b | 0.224 | 21.4 | 55.5 | 349 | 0.399 | 6.80 | 460.512 | 526.3 | 0.001 | 460.01 | 18625.92 |
| 10 | HD 39091 b | 0.228 | 22.4 | 45.9 | 930 | 0.049 | 6.85 | 312.543 | 838.8 | 0.000 | 312.04 | 12188.17 |
| 11 | GJ 328 b | ... | ... | ... | ... | ... | ... | ... | ... | ... | ... | ... |
| 12 | 14 Her b | ... | ... | ... | ... | ... | ... | ... | ... | ... | ... | ... |
| 13 | HD 134987 c | ... | ... | ... | ... | ... | ... | ... | ... | ... | ... | ... |
| 14 | 55 Cnc d | ... | ... | ... | ... | ... | ... | ... | ... | ... | ... | ... |
| 15 | GJ 179 b | ... | ... | ... | ... | ... | ... | ... | ... | ... | ... | ... |
| 16 | HD 87883 b | ... | ... | ... | ... | ... | ... | ... | ... | ... | ... | ... |

Notes: See notes for Table 4.

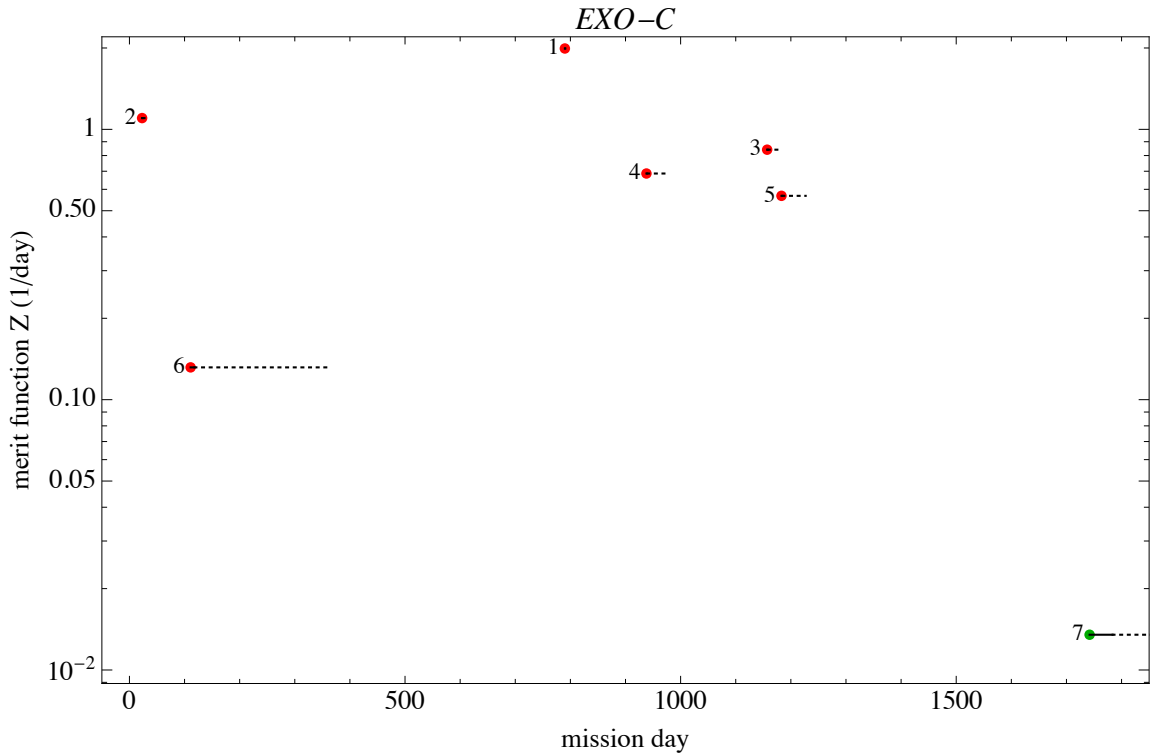

**Figure 11.** Graphical DRM for *EXO-C*. See the caption for Figure 8.

# 7. SUMMARY AND CONCLUSIONS

We have derived a theory for the accuracy of an RV planet's true mass estimated by direct astrometry. The theory involves the RV orbital elements, parameters of the telescope, random values of the unknown inclination angle, and science operations, most notably the scheduling algorithm, merit function, and exposure time. We have derived the signal-to-noise needed to achieve a desired astrometric accuracy—± 10% in the current treatment—using a diffraction-limited, background-dominated telescope in space. Our theory includes node crossings, solar and anti-solar pointing restrictions, photometric and obscurational completeness, exposure-time availability, and image blurring due to orbital motion.

In our definition, this observing program has two goals: to estimate the true planetary mass with accuracy ±10%, and to obtain a coarse spectrum. For mass estimation, the observing strategy is to measure the apparent separation between host star and planet with adequate astrometric accuracy. For the spectroscopy, the strategy is to obtain a spectrum immediately after a productive search image, because the opportunity is perishable.

The information developed in §5–6 reveals the basic nature of the initial observing program to characterize known RV planets. From the illustrative cases in Figures 4–7, we have learned that there are typically only a few, often brief windows in absolute time (Julian days) in which a search observation can be scheduled with high confidence of success. In other words, observations of known RV planets are time-critical and therefore challenging to schedule. Therefore, this paper has developed and illustrated the input information needed by the scheduling algorithm—in this time frame to build diagnostic DRMs and in the future to schedule a real telescope, when it is available.

We have studied the four missions on a basis as equal as possible, for example, assuming a five-year duration for all missions. The sparse and spikey nature of RV-planet availabilities means that the total number of planets characterized may be significantly reduced if the mission duration is reduced.

For the star-shade missions, the exposure times are typically much shorter than the setup time, which in principle means that higher astrometric accuracy can be achieved with little time penalty. Sources of systematic errors may limit the astrometric accuracy achievable in practice.

A corollary of "sparce and spikey" is that potential scheduling conflicts are rare. For the coronagraphs, Figures 10–11 show no conflicts. For the star-shade missions, Figures 8–9 show only two or three conflicts, which should be easily resolvable by small tweaks to the observing times.

We note the mostly short and inconspicuous dashed lines in Figures 8–11, particularly for guaranteed planets (red). These lines show the exposure times for follow-on spectroscopy, beginning immediately after locating the planet in a search image. In most cases, the spectroscopic exposure will be easily accommodated without running into a serious scheduling conflict or major change in the observational constraints.

The star shades out-perform the coronagraphs in this application, by about a factor of three in characterized planets. For both coronagraphs, the input catalog includes 16 RV planets, of which *EXO-C* could possibly observe 10, of which 6 would have a 90% guarantee of success. Of the same 16 planets, *WFIRST-C* could possibly observe 12, of which 9 are guaranteed. For both star-shade missions, the input catalog includes 55 planets, of which *EXO-S* could possibly observe 37, of which 20 are guaranteed. Of the same 55, *WFIRST-S* could possibly observe 45, of which 30 are guaranteed.

We are grateful for Stuart Shaklan's interest in and encouragement of this research. Thanks to Marc Postman's support and guidance. We thank John Krist for helpful input on *WFIRST-C*. Thanks to Mark Marley and Jack Lissauer for inputs on the science program related to known RV planets. This research has made use of the Exoplanet Orbit Database and the Exoplanet Data Explorer at exoplanets.org.